# 7.0 X-rays from Galaxies - Giuseppina Fabbiano

## 7.1 Introduction

Galaxies, of which our own Milky Way is the closest example, are complex systems, composed of stars, gas, dust, non-baryonic (dark) matter and black holes. The deepest images from the *Hubble Space Telescope* show that galaxies populate the universe out to z>8.5 (Ellis et al.2013). *Chandra* has allowed the detection and study of the different components of the X-ray emission galaxies in the local universe out to about 100 Mpc, and of the overall X-ray output of galaxies at higher redshift.

In the X-rays we can uniquely probe the non-thermal emission of stars and the interaction of stellar winds with the surrounding interstellar medium; the end-products of stellar evolution, including both gaseous supernova remnants and compact stellar remnants (white dwarfs, neutron stars); the hot ~10 million degrees gas in the interstellar medium, gaseous outflows and galaxy halos; and a wide range of black holes, from those originating from stellar evolution, to the supermassive black holes at the nuclei of galaxies. These results are setting important constraints on the energy input (feedback) from stellar and nuclear sources during galaxy evolution. During the merging of gas-rich disk galaxies, the X-ray emission is enhanced with increased stellar formation because of both the birth of luminous X-ray binaries and feedback onto the interstellar medium (ISM), which is also enriched of the elements produced by stars and supernovae. Accretion onto the nuclear supermassive black hole is responsible for intense X-ray emission, leading to the formation of a strong Active Galactic Nucleus (AGN), and to galaxy-scale X-rays from the interaction of the AGN photons with the ISM. Hot, extended, X-ray-emitting gas in massive elliptical galaxies points to the presence of massive dark matter halos.

The types of X-ray sources found in galaxies, and the physical processed responsible for the X-ray emission are discussed in detail in other chapters of this book. Here we concentrate on the collective behavior of these sources in their native environment. We will discuss:

(1) The populations of X-ray binaries that compose the larger fraction of the X-ray emission of normal galaxies at energies  > ~2keV (Section 7.2);
(2) The hot gaseous component, prevalent at energies < 1keV, and its evolution in both star-forming and old stellar population galaxies (Section 7.3);
(3) The discovery of low-luminosity 'hidden' active nuclear sources in normal galaxies, that are expanding the range of nuclear back holes down to 'intermediate' masses, bridging the gap between stellar black holes (Mass up to ~100$M_\odot$) and supermassive nuclear black holes (M~ $10^8$-$10^{10}$ $M_\odot$; Section 7.4); and finally
(4) The interaction of supermassive back holes with their host galaxy, which pose constraints both on nuclear feedback, and on our understanding of the nature of active galactic nuclei (AGNs; Section 7.5).

The results discussed in this chapter could not have been achieved without the sub-arcsecond resolution of the *Chandra* X-ray telescope, and the ability of the ACIS detector to record simultaneously position, energy, and time of the incoming photons (Chapter 2). However, the small collecting area of *Chandra* requires very long exposures to pursue this science. All of this makes an excellent scientific case for a large-area next generation X-ray telescope that retains the angular resolution of *Chandra*, and is equipped with focal plane instruments capable of providing higher resolution in the spatial, spectral and time domains.

The following discussion also demonstrates the importance of multi-wavelength observations of comparable quality, provided by *HST*, *Spitzer* in space and the *JVLA* and *ALMA* from the ground, for achieving a full understanding of the processes in action. It is important that this multi-wavelength capability be retained by the astronomical community in the years to come, if we do not want the type of science possible in the *NASA Great Observatory* era to became an example of an irretrievable golden past (Section 7.6).

**7.2 X-ray Binary Populations**

X-ray astronomy began with the unexpected discovery of a very luminous source, Sco X-1 (Giacconi et al.1962). Sco X-1 was the first Galactic X-ray binary (XRB) ever to be observed. The first X-ray survey of the sky with the NASA *Uhuru* satellite (also conceived and led by Giacconi and collaborators) showed that XRBs are the most common luminous X-ray sources in the Milky Way. XRBs are binary systems composed of an evolved stellar remnant (Neutron Star –NS, Black Hole – BH, or White Dwarf -WD), and a stellar companion. The X-rays are produced by the gravitational accretion of the atmosphere of the companion onto the compact remnant (for reviews on XRBs, see 'Xray Binaries', eds. Lewin, van Paradijs & van den Heuvel 1995; and 'Compact stellar X-ray sources', eds. Lewin & van der Klis 2006).

If the companion is a massive star (mass > 10 $M_\odot$), the XRB is called a High Mass X-ray Binary (HMXB); HMXBs are short-lived X-ray sources, with lifetimes ~ 10 Million years, regulated by the evolution of the massive companion, and therefore are associated with young stellar populations. If the companion is a low-mass star (mass ≤ 1$M_\odot$), the XRB is called Low Mass X-ray Binary (LMXB). There are two types of LMXB: (1) those resulting from the evolution of native stellar binary systems in the parent galaxy stellar field (field-LMXBs; see Verbunt & van den Heuvel 1995); and (2) those formed from dynamical interactions in Globular Clusters (GC-LMXBs; Clark 1975; Grindlay 1984). Field-LMXBs evolve at a slow pace, with lifetimes of ~$10^8$-$10^9$ yr, due to the long time needed to produce a close binary and start the accretion process, so these LMXBs are generally old systems. Instead, GC-LMXBs can be formed virtually at any time, and some of them may be short-lived systems (e.g., the NS-WD ultra-compact binaries, Bildsten & Deloye 2004).

Ever since the discovery of Sco X-1, the study of Galactic XRBs has continued to be a thriving area of research. This body of work has investigated the physical processes related to the gravitational accretion onto a compact object in a binary system, and has set constraints on the mass and nature of the compact stellar remnants. In the late 1970's and 1980's, following the advent of imaging X-ray astronomy with the *Einstein Observatory*, XRBs were detected in a number of nearby galaxies (see review, Fabbiano 1989). While observations with other X-ray telescopes (*ROSAT*, *ASCA*, *XMM-Newton*) have contributed to the study of extra-galactic XRBs, the sub-arcsecond resolution of the *Chandra X-ray Observatory* (Weisskopf et al. 2000) has revolutionized this field (see review Fabbiano 2006). With *Chandra*, populations of luminous point-like sources (typically $L_X > 1 \times 10^{37}$ erg s$^{-1}$ in the Chandra energy band ~0.5-7 keV) have been detected in all galaxies within 50 Mpc of the Milky Way and even farther away (Fig. 7-1).

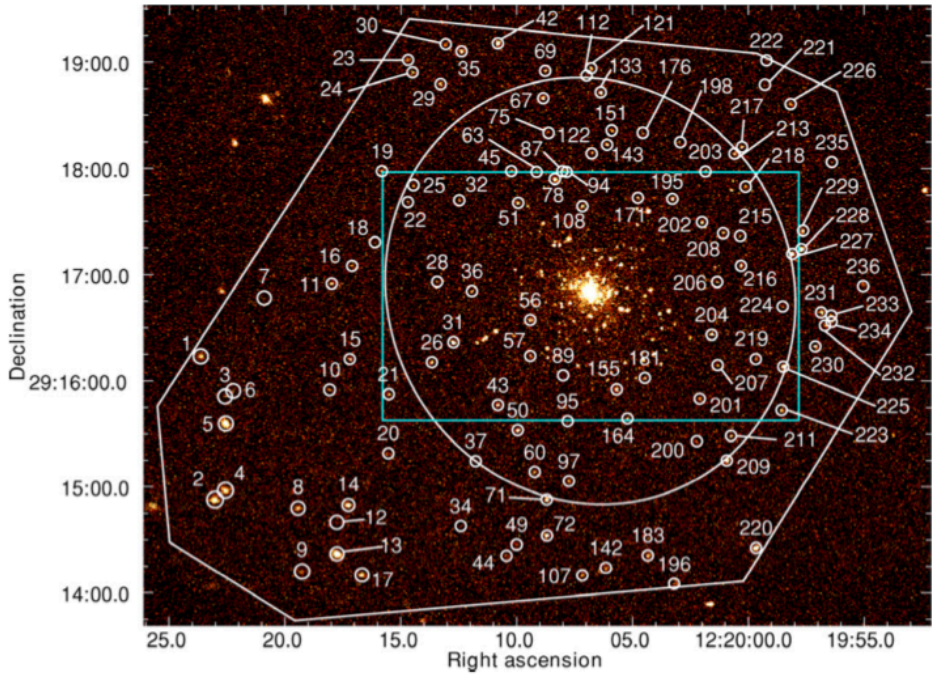

Fig. 7-1 – The LMXB population of the elliptical galaxy NGC 4278 (Brassington et al. 2009). The top panel shows the sources detected at larger radii, within the footprint of the Chandra fields (polygon). The ellipse in the top panel represents the D25 isophote of the optical stellar light. The left bottom panel expands the region within the blue rectangle in the top panel, and the right bottom panel expands the region within the blue rectangle in the left botton panel. NGC 4278 was observed with *Chandra* ACIS-S in six separate pointings, resulting in a co-added exposure of 458 ks. From this deep observation, 236 sources have been detected within the region overlapped by all observations, 180 of which lie within the D25 ellipse of the galaxy. These 236 sources (of which only 29 are expected to be background AGNs) range in $L_X$ from $3.5 \times 10^{36}$ erg s$^{-1}$ (with 3σ upper limit ≤$1 \times 10^{37}$ erg s$^{-1}$) to ~$2 \times 10^{40}$ erg s$^{-1}$. They have X-ray colors consistent with those of Galactic LMXBs. Of these sources, 103 are found to vary long-term of various amounts, including 13 transient candidates (>5 ratio variability).

A large body of work convincingly associates these sources with XRBs (we refer the reader to Fabbiano 2006). In particular, the X-ray colors of the majority of these sources are consistent with the spectra of Galactic LMXBs and HMXBs. Classes of softer emission sources (Super Soft Sources and Quasi-Soft Sources), possibly associated with nuclear burning white dwarf binaries, were also detected. Monitoring observations of some of these systems have uncovered the widespread source variability characteristic of XRBs, pointing to compact accreting objects. Spectral variability patterns consistent with those of BH LMXBs were found in sources detected in the nearby elliptical galaxy NGC 3379 (Brassington et al. 2008). In this galaxy, and in NGC 4478, the spectra of most luminous LMXBs are consistent with the emission of accretion disks of black holes (BHs) with masses of 5-20 $M_\odot$ (Brassington et al. 2010; Fabbiano et al. 2010). These masses are in the range of those measured in Milky Way BH binaries (Remillard and McClintock 2006).

*Chandra* observations of galaxies provide a new tool for constraining the formation and evolution of XRBs. Given that the XRBs in a given galaxy are all at the same distance, with *Chandra* we can derive their relative luminosities accurately (which is not the case for Galactic XRBs); the absolute luminosities will be affected by the uncertainty on the distance of the parent galaxy. Moreover, we can correlate the properties of these sources with those of the parent stellar population. By detecting an ever growing sample of XRBs we are also sensitive to extreme objects, that may be missing in the Milky Way, such as the Ultra-Luminous X-ray sources (ULXs) that emit in excess of the Eddington luminosity for a NS or ~10 $M_\odot$ stellar BH system (>$10^{39}$ erg s$^{-1}$). ULXs have been suggested to be intermediate mass BHs, bridging the gap between the supermassive BHs of AGNs and the stellar BHs found in some XRBs, although now most systems are understood to be young luminous HMXBs (see Fabbiano 1989, 2006; see also Section 7.2.1.2).

**7.2.1 The XRB X-ray Luminosity Functions (XLFs) and Scaling Laws in the Near Universe**

The XRB populations of galaxies can be characterized by means of their X-ray luminosity function (XLF). In differential form, the XLF gives the number of XRBs detected at a given X-ray luminosity; in integral form, the number of XRBs detected above a given X-ray luminosity (Fig. 7-2). The XLFs can be parameterized in terms of their normalization (i.e. the number of X-ray sources detected in a given galaxy), and shape (i.e. the distribution of theses sources in function of X-ray luminosity; functional slopes and break $L_X$). Both parameters reflect the composition of the XRB populations (see the review of Fabbiano 2006 and refs. therein).

The XLF normalization measures the total number of XRBs in a given galaxy, and thus the total X-ray luminosity of the galaxy, at least at energies > 2 keV, where XRBs dominate the X-ray emission of normal galaxies. Beginning from the earlier studies of normal (non-active) galaxies in X-rays with the *Einstein Observatory*, it has been evident that both total stellar mass and star formation rate are important factors in determining the total number of XRBs. Total stellar mass is the prevalent scaling factor for old stellar population, i.e. elliptical galaxies and bulges, while star formation rate drives XRB production in the young stellar populations of spiral disks and irregular galaxies (see review Fabbiano 1989, 2006 and refs. therein).

The shape of the XLF measures the relative number of low- and high-luminosity XRBs in a given population, and therefore provides the means to constrain the luminosity evolution of these sources (see e.g., Fragos 2013b). Several monitoring studies of individual galaxies show that although individual XRBs vary strongly in their X-ray output, this variability does not affect the general shape of the XLF (Zezas et al. 2007; Fridriksson et al. 2008; Mineo et al. 2014c).

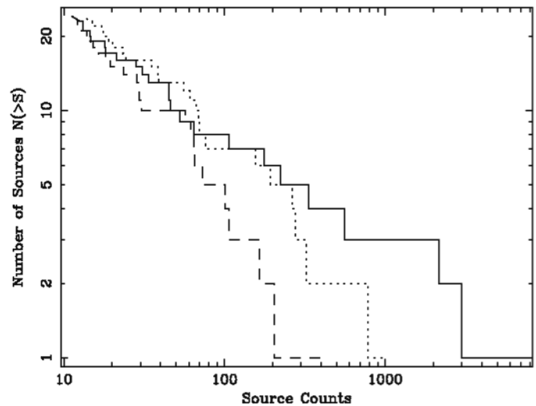

Fig 7-2 – Integral XLFs of sources in M81: solid – arms; dotted – moving towards the disk; dashed – disk (Swartz et al. 2003)

*7.2.1.1 Old Stellar Populations*

In old stellar systems, such as elliptical galaxies, the XRB population is composed of low mass stars accreting onto a compact remnant (LMXBs). As mentioned earlier in Section 7.2, two evolution channels are possible: (1) the evolution of a native X-ray binary composed of a high mass star (now evolved into a compact remnant) and a long-lived low mass stellar companion, or (2) dynamical binary formation in GCs. Because of its efficiency, the latter has been argued to be the main formation mechanism for LMXB (Clark 1975). *Chandra* observations of nearby elliptical galaxies suggest that both formation channels are important. In particular, the normalization of the XLFs of these XRB populations is dependent largely on the integrated stellar mass of the parent galaxy and to a minor, but significant, extent, on the GC specific frequency (number of GCs per unit stellar mass; see review Fabbiano 2006; Kim et al. 2009; Boroson et al. 2011). These dependencies suggest that both LMXBs from the general stellar population and from GCs are involved.

The shape of the LMXB XLF also reflects the nature and evolution of the LMXBs. Based on joint *Chandra* and *Hubble* surveys of nearby elliptical galaxies, it has been possible to disentangle the XLFs of the LMXBs detected in the galaxy stellar field and those associated with GCs. While earlier reports suggested that the XLFs of the GC-LMXB and field-LMXB did not differ (see Fabbiano 2006), later work with deeper data have demonstrated that the GC-LMXB XLF is flatter than that of the field-LMXB. This difference was first suggested only for the lower luminosity portion of the XLF, $L_X < 5 \times 10^{37}$ erg/s (Kim et al. 2006). More recent work that combines the LMXB populations of several galaxies has concluded that the GC-LMXB XLF is overall flatter (Lehmer et al. 2014; Peacock & Zepf 2016).

Both the metallicity and the age of the stellar populations can affect the XLF. Metal rich, red, GCs host on average ~3 times more LMXBs than metal poor, blue GCs, affecting the normalization of the relative XLFs. However, no significant difference in XLF shape exists between LMXBs associated with metal rich or metal poor clusters (Kim et al. 2013). Instead, old stellar populations of relatively younger stellar ages have been associated with a relative excess of luminous LMXBs, resulting in flatter XLF slopes (Kim & Fabbiano 2010; Zhang, Gilfanov & Bogdan 2012). Further work that disentangled GC-XLF and field-XLF has clearly demonstrated that this increased number of luminous sources is a feature of the field population, which is also predicted by LMXB evolution models (Lehmer et al. 2014; Fragos 2013b).

*7.2.1.2 Young and Mixed Stellar Populations*

In the young stellar populations of actively star-forming galaxies, such as late-type spirals and irregulars in the Hubble sequence, the XRB population is dominated by young, massive, binary systems with a compact stellar remnant (NS or BH) and a high mass stellar donor companion. These HMXBs are believed to be the product of the evolution of native binary systems, and have relatively short lifetimes ($\sim 10^7$ yrs), due to the rapid evolution of the massive donor star. This is the less massive of the original pair, the more massive having already evolved into the compact remnant.

The number of XRBs in star forming galaxies (XLF normalization) is proportional to the star formation rate (SFR) of the stellar population of the galaxy. The shape of the XLF is a flat power-law extending to high X-ray luminosities. In galaxies with very high SFR, such as colliding/merging galaxies (e.g., the Antennae galaxy, Zezas et al. 2007; NGC2207/IC2163, Mineo et al. 2014c), the XLF extends up to luminosities of $\sim 10^{40}$ erg/s, including significant numbers of ULXs (sources detected at luminosities greater than $10^{39}$ erg/s). While it has been suggested that ULXs may host intermediate mass BHs (see Fabbiano 2006), the continuity of these XLFs is part of the evidence for ULXs belonging to the HMXB population and being powered by super-Eddington accretion onto stellar mass BHs (King et al. 2001; see also Section 7.2.3). Even a NS can power an ULX, as shown by the M82 pulsar ULX (Bachetti et al. 2014, Fragos et al. 2015).

Studies of individual galaxies with complex stellar populations, such as spiral and late-type galaxies, demonstrate that the characteristics of the XLF closely correlate with the local stellar population (see review, Fabbiano 2006). For example, in the interacting starburst galaxy NGC2207/IC 2163, the number of ULXs (28 in all) correlates with the local SFR (Mineo et al. 2013, 2014c), again linking these sources to the normal stellar binary population. A new study of M51 exemplifies the link between the properties of the stellar population (SFR and mass-weighted stellar age) and the XLF (Lehmer et al. 2017). Differences are also reported in the XLF and source properties of the bulge and ring XRB populations in the ring galaxy NGC 1291 (Luo et al. 2012), reflecting the different ages of these stellar populations.

A mix of stellar populations is normal in star forming galaxies. Therefore their XRB populations should include both HMXBs and LMXBs, as is clearly the case in the Milky Way. Under this assumption, Lehmer et al. (2010) parameterized the total XRB X-ray luminosity of a sample of nearby luminous IR galaxies as a function of both stellar mass (for the LMXB population) and SFR (for the HMXB population), proposing the relation $L_X^{gal} = \alpha M_\star + \beta SFR$, where $\alpha = (9.05 \pm 0.37) \times 10^{28}$ erg s$^{-1}$ M$_\odot^{-1}$ and $\beta = (1.62 \pm 0.22) \times 10^{39}$ erg s$^{-1}$ (M$_\odot$ yr$^{-1}$)$^{-1}$. These authors point out that HMXBs dominate the galaxy-wide X-ray emission for galaxies with specific star formation rate SFR/M$_\star \geq 5.9 \times 10^{-11}$ yr$^{-1}$. This baseline near-universe-based relationship can be used to investigate the evolution of XRB populations at different redshifts.

**7.2.2 The Redshift Evolution of the XRB Emission**

With the advent of *Chandra* surveys [e.g. the ever-deepening Deep Fields (Brandt et al. 2001) and the ever-widening wide area medium depth COSMOS survey (Elvis et al. 2009; Civano 2012)], observations of normal galaxies have been extended from the near universe out to z∼5, allowing studies of the redshift evolution of the XRB population. Multi-wavelength coverage of these surveys has provided measurements of both the total stellar mass (M∗) and the SFR of the observed galaxies, leading to the parameterization of the integrated XRB emission of galaxies in function of these parameters and of redshift.

Both Lehmer et al. (2008), from the stacking of galaxies covered by the *Chandra* Deep Fields (up to 2Ms *Chandra* exposure), and Mineo et al. (2014b), using a sample composed of near universe, high luminosity IR galaxies, and galaxies detected in the *Chandra* Deep Fields, find a linear relation between total $L_X$ and SFR, and conclude that the X-ray emission can be used as a robust indicator of star formation activity out to z ≈ 1.4. Subsequently, Lehmer et al. (2016), using the deeper 6 Ms CDF-S, find that simple SFR scaling is insufficient for characterizing the average X-ray emission at all redshifts, and establish a scaling relation involving both SFR and stellar mass. Aird, Coil & Georgakakis (2017), using several deep surveys, propose in addition a power-law dependence of the SFR. More recently, Fornasini et al. (2018) using the stacked luminosities of ∼75,000 star-forming galaxies in the COSMOS survey with 0.1<z<5, confirm the functional dependence of the Aird et al. relationship: $L_X = \alpha(1+z)^\gamma M_\ast + \beta(1+z)^\delta SFR^\theta$ and find best-fit values of the parameters: log($\alpha$) = 29.98 ± 0.12, $\gamma$ = 0.62± 0.64, log($\beta$) = 39.78 ± 0.12, $\delta$ < 0.2, $\theta$ = 0.84±0.08.

It is likely that future studies will provide more complex relationships for the z-dependence of the X-ray emission of XRB populations. For example, simulations of XRB populations and their XLFs have pointed out that metallicity, and initial stellar mass function (IMF) slope are also important factors in XRB evolution and in the X-ray output (e.g., Fragos et al. 2013a, b; Tremmel et al. 2013). These conclusions await a thorough observational verification.

Careful characterization of the redshift evolution of the XRB population is important not only for understanding the history of XRBs, but also for studying the other components of the cumulative X-ray emission of galaxies, in particular the occurrence and properties of low-brightness hot halos (Section 7.3) and nuclear emission (Section 7.4). In principle, once the parameters of the XLFs and their evolution are well constrained observationally, and well reproduced by models, X-ray observations may provide a way to constrain galaxy evolution. XRB energy input may be important both in the early universe (Fragos et al. 2013b; Lehmer et al. 2016), and perhaps in later galaxy feedback.

### 7.2.3 The Spatial distributions of the XRBs

Since the earliest detections of normal galaxies in X-rays, it has been known that the spatial distribution of the hard (>2 keV) XRB-dominated X-ray emission tends to follow that of the integrated stellar light, i.e. their parent stellar population (see Fabbiano 1989 and refs. therein). *Chandra* observations have clearly demonstrated this general association with the detection of individual XRBs in galaxies (see Fabbiano 2006 and refs. therein). As discussed in Section 7.2.1.2, in star forming galaxies with complex stellar populations, *Chandra* imaging has allowed the spatially resolved investigation of the connection of XRB populations with the local underlying stellar population. These spatial associations have provided interesting constraints on the nature and evolution of XRBs in both young and old stellar populations.

The definitive associations of very luminous ULXs with intensely star-forming regions, (e.g., in the Antennae, Fabbiano et al. 2001; the Cartwheel galaxy, Wolter & Trinchieri 2004; and many other cases) support the conclusion that ULXs belong to these young stellar populations (see also Section 7.2.1.2). Given the typical stellar masses of these populations, ULXs are likely to be young binary systems with super-Eddington accretion (King et al. 2001; Soria et al. 2009). A comparison of the positions of ULXs with the Sloan survey database (Swartz et al.2009) concludes that ULXs tend to be associated with OB associations, rather than with super-star-clusters, consistent with the results on the Antennae galaxies (Zezas et al. 2002). Interestingly, the colors of the stellar regions associated with ULXs tend to be redder of those of HII regions. If this reddening is not due to localized absorption, the redder colors would be consistent with an aging of the stellar population commensurable with the evolution time of a massive X-ray binary.

Observations of stellar populations too young for HMXB formation validate the evolutionary models of massive binary systems into HMXBs. This evolution requires a few ~ $10^6$-$10^7$ years. Therefore, although the normalization of the XLF is proportional to the SFR, more properly it should be said that the number of HMXBs represents the star formation that took place in a galaxy 5-60~Myr ago. Shtykovskiy & Gilfanov (2007) compared the distributions of XRBs and HII regions in the spiral arms of M51 and suggest that the distribution of HMXBs is not as peaked as that of the HII regions, reflecting earlier occurrences of star formation that have been left behind by the spiral compression wave. In the Large Magellanic Cloud (LMC), our nearest star-forming galaxy, Antoniou and Zezas (2016) find that the HMXBs are present in regions with star formation bursts of ~6-25 Myr age; in the Small Magellanic Cloud (SMC), instead, the HMXB population peaks at later ages (~25-60 Myr; see also Antoniou et al.2010). Antoniou & Zezas (2016) also find that the formation efficiency of HMXBs in the LMC is ~17 times lower than that in the SMC and attribute this difference primarily in the different age and metallicity of the HMXB populations in the two galaxies. The SMC formation age is consistent with the findings of Williams et al. (2013) of 40–55 Myr formation age for the HMXBs in NGC 300 and NGC 2403.

In elliptical galaxies, several studies of the radial distributions of detected X-ray sources have been pursued in attempts to associate the LMXBs with either the stellar field or with the GC systems. These studies led to inconclusive results, in part due to the relatively poor statistics of the LMXB populations and to incomplete mapping of the GC systems (see Fabbiano 2006). More recently, Zhang et al. (2013) investigated the cumulative radial distribution of LMXB in 20 early – type galaxies observed with *Chandra*, and reported that sources more luminous than $\sim 5 \times 10^{38}$ erg/s follow the distribution of the stellar light, while observing an over-density of fainter sources, out to at least $\sim 10 r_e$ ($r_e$ is the effective radius). They proposed that the extended LMXB halos may be comprised of both low-mass X-ray binaries (LMXBs) located in GCs, which are known to have a wider distribution than the stellar light, and of neutron star LMXBs kicked out of the main body of the parent galaxy by supernova explosions.

The complete spatial study of the rich LMXB and GC populations of NGC 4649, a giant E in the Virgo cluster, was made possible by the coordinated complete deep coverage of these populations with *Chandra* and *HST* surveys. These observations resulted in complete samples of both LMXBs and GCs and in reliable identifications of X-ray sources with GC counterparts (Strader et al. 2012; Luo et al. 2013). Using these rich data sets, Mineo et al. (2014a) showed that GC-LMXBs have the same radial distribution as the parent red and blue GCs. The radial profile of field LMXBs follows the V-band profile within the $D_{25}$[1] of NGC 4649, consistent with an origin of these sources from the evolution of native binaries in the stellar field. Mineo et al. (2014a) also suggest a possible excess of LMXBs in the field population at large radii that may be consistent with the report of Zhang et al. (2013).

A different approach to the study of the spatial distribution of LMXB and GCs is provided by the analysis of the 2-dimensional distributions of LMXBs and GCs on the plane of the sky, and on the detection and characterization of significant localized discrepancies from the azimuthally smooth distributions conforming to the radial profiles of these objects (Bonfini et al. 2012; D'Abrusco et al. 2013). This technique has revealed significant anisotropies, suggesting streamers from disrupted and accreted dwarf companions in virtually all the galaxies thus analyzed (D'Abrusco et al. 2013; 2014a, b; 2015).

In both NGC 4649 (D'Abrusco et al. 2014a – Fig. 7-4) and NGC 4278 (D'Abrusco et al. 2014b), arc-like distributions of GCs are associated with similar over-densities of X-ray sources. In NGC 4649 the GC-LMXBs follow the anisotropy of red GCs, where most of them reside. However, a significant over-density of (high-luminosity) field LMXBs is also present to the south of the GC arc, suggesting that these LMXBs may be the remnants of star formation connected with a merger event. Alternatively, they may have been ejected from the parent red GCs, if the bulk motion of these clusters is significantly affected by dynamical friction. These sources occur at relatively large galactocentric radii and certainly contribute to the excess of field LMXB reported by Mineo et al. (2014a).

---

[1] $D_{25}$ is the apparent major isophotal diameter, measured at or reduced to the surface brightness level 25.0 B-mag per square arcsecond, from the Third Reference Catalog of Bright Galaxies, de Vaucouleurs et al 1991, Springer-Verlag

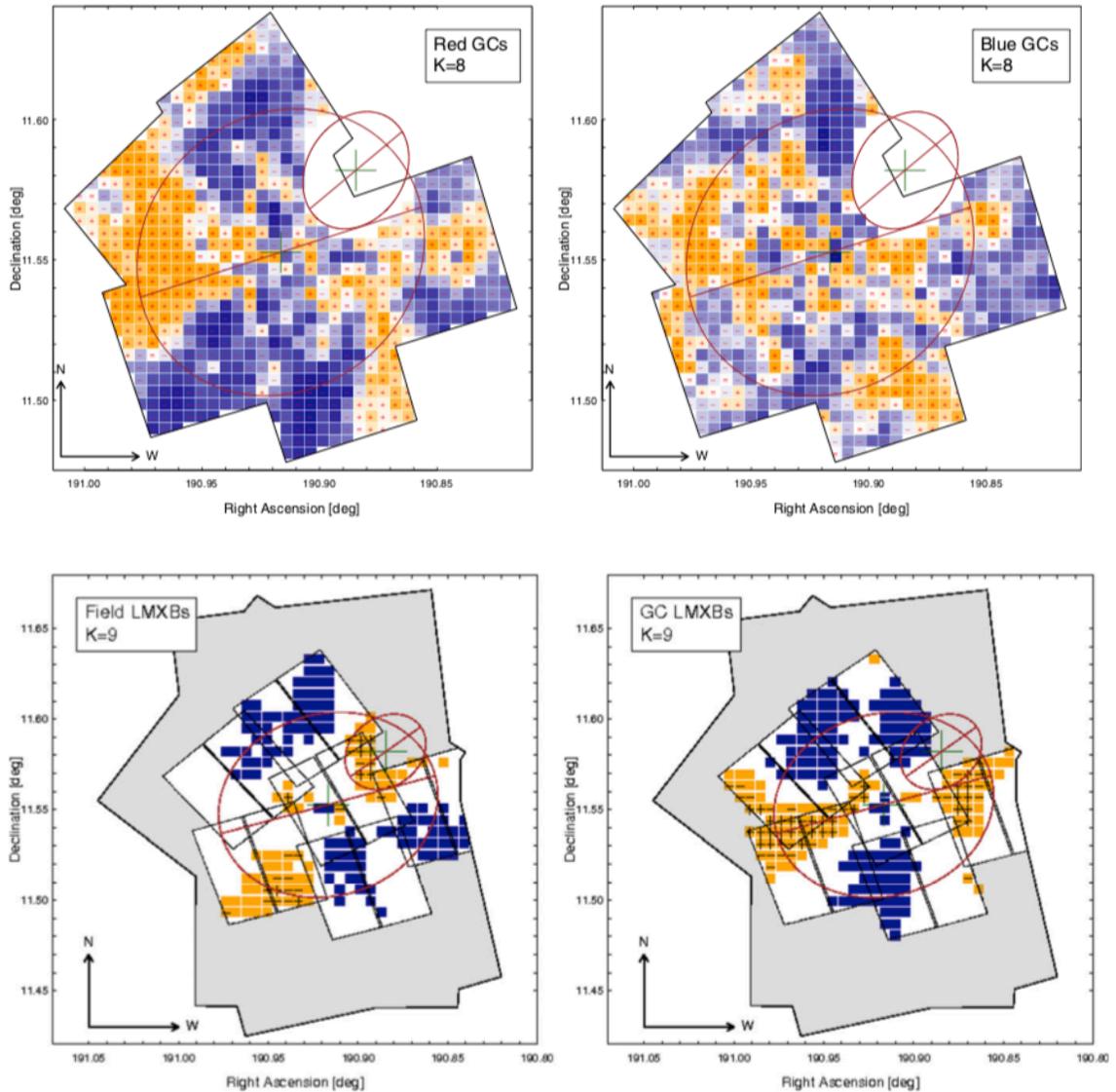

Fig. 7-3 – From D'Abrusco et al. (2014a). Residuals of the 2-dimensional distributions of GCs in NGC4649 (top, with the HST fields footprint) and LMXBs (bottom, with both HST and Chandra footprints), relative to an azimuthally smooth distribution. Yellow are positive residuals, blue are negative. See D'Abrusco et al. for details. In all panels, the larger red ellipse is the D25 of NGC 4649, and the smaller red ellipse is the D25 of a nearby spiral galaxy.

### 7.3 Hot ISM and Halos

The first attempts to study the hot ISM systematically in normal galaxies were made with the *Einstein Observatory*: hot outflows and winds were detected in nearby starburst galaxies (e.g. NGC 253 and M 82); extended hot gaseous components were found in some giant elliptical galaxies in the Virgo cluster. But the lack of good angular resolution in *Einstein* (~1 arcmin) and all other pre-*Chandra* X-ray observatories left room for speculation, since the gaseous emission could not be cleanly separated from other sources, primarily the XRB populations that we discussed in Section 7.2. For example, the widespread presence of hot gaseous emission in elliptical galaxies was strongly debated (see Fabbiano 1989). Even when the spectral separation of these components was later attempted with the advent of X-ray CCDs in the Japanese X-ray satellite *ASCA*, some results were markedly strange, such as the very low metal abundances found in these gaseous components, in both star-forming and elliptical galaxies. These low abundances were in marked contrast with all expectations of chemical evolution (see e.g., the review of the history of these studies for elliptical galaxies in Fabbiano 2012).

Because of its joint sub-arcsecond spatial resolution and spectral capabilities, the *Chandra* telescope with the ACIS CCD camera is well suited to the study of the extended hot interstellar medium (ISM) of galaxies. With the advent of *Chandra* most issues and controversies from previous studies were resolved, and a new discovery space was opened to astronomers. Below we discuss some of these new developments for both the hot ISM of star forming galaxies and that of early-type galaxies (ETGs, including both elliptical and S0s), where it is sometimes called a 'hot halo'. The hot halos of ETGs are also discussed in Chapter 9, especially from the point of view of the physical processes governing these halos.

#### 7.3.1 The Hot ISM of Star-Forming Galaxies and Mergers

In star-forming galaxies, supernova explosions and winds from massive stars heat the ISM and enrich it with the elements shed by the stars. X-ray observations probe both the physical and the chemical properties of this hot component. In starburst regions, if the energy input into the ISM is enough to counterbalance or even surpass the pull of gravity, the hot ISM will expand above the disk of the galaxy, and in the most intense starbursts escape in the form of a hot wind (Chevalier & Clegg 1985; Heckman et al. 1990). These hot winds and outflows were first detected with the *Einstein Observatory* (e.g. in NGC 253 and M82, Fabbiano 1988) extending a few to ~10 kpc out of the plane of the galaxy. Outflows are frequently associated with optical line emission (Heckman et al. 1990) and may be common in star-forming galaxies at high redshift (Martin et al. 2012).

*7.3.1.1 Physical Evolution of the hot ISM*

With *Chandra* observations, the connection between the heating of the ISM and star formation activity has been definitively established. Systematic *Chandra* studies of the soft, $kT$~0.2-0.7 keV (T~ 2-8 million K), diffuse emission of nearby galaxies show that the higher the star formation rate of a galaxy (and therefore the supernova rate), the higher is the luminosity of the hot ISM (Mineo et al. 2012). Observations of edge-on galaxies in the near universe also show that soft emission extending outside of the galaxy plane is common in intensely star-forming galaxies. The presence of extra-planar X-ray emission is always associated with extra-planar optical line emission of similar vertical extent (Strickland et al. 2000; 2004).

Joint *Chandra* and optical studies of the nuclear outflows of NGC 253 (Strickland et al. 2000) advance the picture of the conical outflow of a hot (>2 keV) superwind, fueled by the starburst. The soft (<2 keV) X-rays, and the optical Hα line emission would result from the interaction of this energetic wind with the swept up and entrained gas from the denser ambient interstellar medium. Modeling of *Chandra* and *XMM-Newton* observations of M82 (Strickland & Heckman 2009) suggests that the plasma within the starburst region (inside the circle in Fig. 7-4) has a temperature $T \sim$ 30–80 million K ($kT \sim$ 2-7 keV), a mass-flow rate out of the starburst region of 1.4–3.6 $M_\odot$ yr$^{-1}$ and a terminal wind velocity $v_\infty \sim$ 1410–2240 km s$^{-1}$. This velocity surpasses both the escape velocity from M82 ($v_{esc}$ < 460 km s$^{-1}$) and the velocity of the Hα emitting clumps and filaments entrained in M82's wind ($v_{H\alpha} \sim$ 600 km s$^{-1}$).

Modeling of these data led Strickland & Heckman (2009) to estimate relatively high supernova and stellar wind thermalization efficiencies (30% < ε < 100%). These high efficiencies are in contrast with the conclusions of Mineo et al. (2012) and Richings et al. (2010) that on average only 5% of the mechanical energy of supernovae is converted into thermal energy of the ISM. The latter authors, however, only considered the thermal energy corresponding to the soft (<2 keV) ISM detected in their galaxy samples.

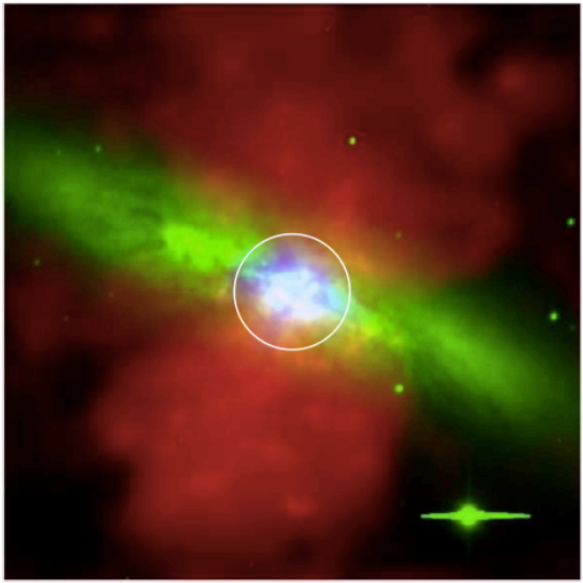

Fig. 7-4 – From Strickland & Heckman 2009 – Composite image of M82. Red, soft outflowing wind: 0.3-2.8 keV, blue: nuclear starburst, 3-7 keV, and green: star light optical emission. The X-ray data are from Chandra ACIS, adaptively smoothed after point source subtraction and interpolation. The circle identifies the 500 pc radius region around the nucleus.

Interacting and merging galaxies provide a local laboratory where astronomers can easily observe phenomena that occur in the deeper universe. There, merging is common and may be an important step in the evolution of galaxies (e.g., Navarro, Frenk, & White 1995). Major mergers of similar mass spiral galaxies may give rise to an elliptical galaxy (Toomre & Toomre 1972). Galaxy interaction and merging triggers enhanced star formation, which also means an increased injection of energy in the ISM. This may cause galaxy-size outflows, and large-scale extended structures and hot halos (Fig. 7-5).

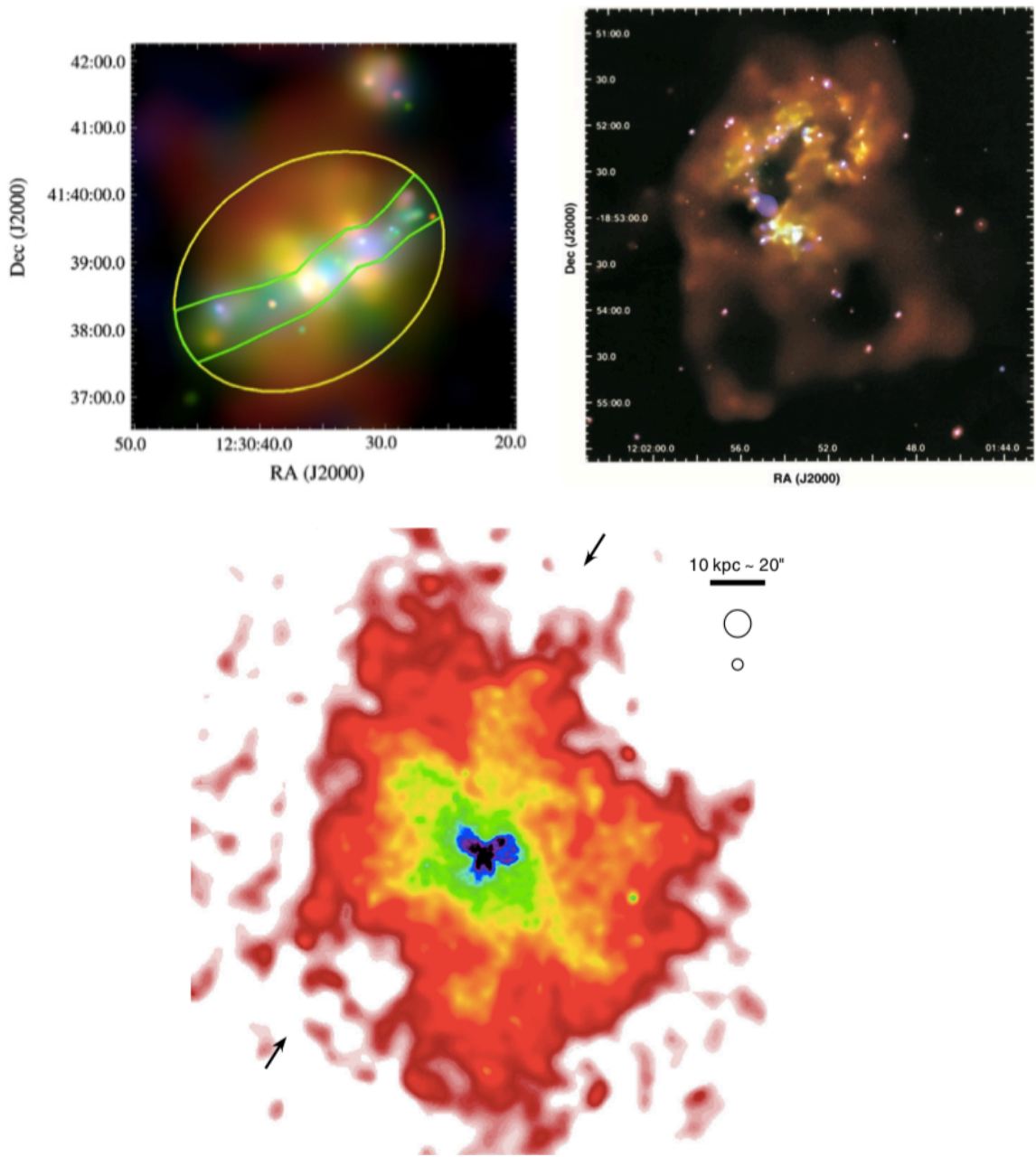

Fig. 7-5 -Three examples of *Chandra* images of large scale structures in the hot gas associated with interacting and merging galaxies. N is up, E to the left in all these images. Top left, the hot ISM of the interacting galaxy NGC 4490 (Richings et al. 2010), with the companion galaxy NGC 4485 identifiable with the clump of emission to the north. Point sources (XRBs) were subtracted from this image that shows four outflow regions from the galaxy plane and a fainter (red) halo extending ~7.5 kpc out of the plane; the green lines give the outline of the plane and the yellow ellipse that of the halo. Top right, the image of the Antennae galaxies showing the two giant loops (10 kpc across) to the south (Fabbiano et al. 2004). Bottom, the image of NGC6240 (blue higher intensity, red lower intensity, see Nardini et al. 2013) showing the loops and filaments in the hot ISM embedded in a giant hot halo (80 x 100 kpc).

As shown in Fig. 7-5 (top-left), a large-scale halo is detected in the interacting system NGC 4490 / NGC 4485 (Richings et al. 2010), extending ~7.5 kpc ouside the galaxy plane. This halo is not consistent with a freely expanding adiabatically cooling wind, as may occur in M82, so it could be at least partially gravitationally bound in the dark matter potential of the galaxy.

In the merging pair the Antennae galaxies (NGC 4038/39; Fig. 7-5, top-right), besides widespread intense hot ISM emission throughout the stellar disks, two giant loops of hot gas (~10 kpc across, kT~ 0.3 keV) are seen extending to the south of the merging stellar disks. A cooler ~0.2 keV low surface brightness hot halo, extends out to ~18 kpc. These features may be related to superwinds from the starburst in the Antennae or result from the merger hydrodynamics. Their long cooling times (~1 Gyr) suggest that they may persist to form the hot X-ray halo of the emerging elliptical galaxy (Fabbiano et al. 2004).

NGC 6240 (Fig. 7-5, bottom), a more advanced merger than the Antennae, displays several loops of hot X-ray emitting gas, spatially coincident with Hα filaments and a very extended and luminous X-ray halo with projected physical size of ~110 × 80 kpc (Nardini et al. 2013). This halo could be pre-existent to the starburst. In a few cases of massive spiral galaxies there have been reports of extended static hot coronae, trapped in the galaxy potential (Bogdán et al. 2013).

*7.3.1.1 Chemical Evolution of the hot ISM*

*Chandra* observations have resolved a long-standing issue with the measurements of metal abundances in the hot ISM from spectral fitting of X-ray data. Pre-*Chandra* observations of star-forming galaxies suggested that the hot ISM had very low (sub-solar) metal abundances. This is odd, given that the hot ISM should be enriched by the supernova ejecta and stellar winds in the star-forming regions (e.g., Weaver et al. 2000). In the Antennae, for example, *ASCA* spectra taken with CCDs with similar energy resolution to those in the *Chandra* ACIS, suggested an overall extremely low abundance of heavy elements (~0.1 the solar value; Sansom et al. 1996).

The *Chandra* ACIS observations of the Antennae provided the opportunity for a direct comparison with the *ASCA* results. Baldi et al. (2006a) demonstrated that the puzzling sub-solar abundances derived from the *ASCA* spectra were the result of mixing the signals from regions of the hot ISM of different spectral properties. They analyzed the entire *Chandra* emission from the Antennae as a single source, as it would have been seen by *ASCA,* given its lack of angular resolution, and were able to reproduce the sub-solar Sansom et al.'s results. Their detailed analysis of the *Chandra* data set of the Antennae, separating emission regions of different intensity and morphology, instead returned much higher – several times solar - abundance values, differing in different regions. *Chandra* studies of other starburst galaxies, analyzing separate emission regions within a given hot ISM complex, also returned higher metal abundances (e.g., Martin et al. 2002; Richings et al. 2010; Nardini et al. 2013). Mixing contributions of different temperature regions led to overlapping line emission that formed a pseudo continuum seen by ASCA.

In the Antennae, the ratios for several elemental abundances are consistent with those expected for SN II, the type of supernova resulting from the evolution of the massive stars found in starburst regions, but not SNIa. (Baldi et al. 2006b, Fig. 7-6). A similar result was reported for NGC 4490 (Richings et al. 2010).

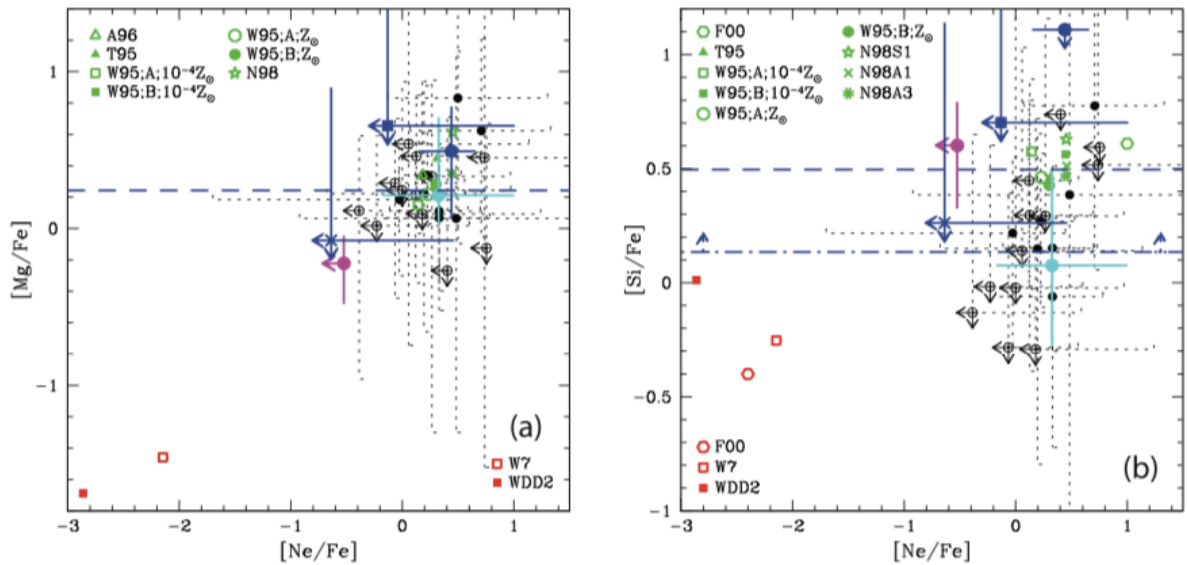

Fig. 7-6 - Elemental abundance ratios from distinct regions of the hot ISM of the Antennae (points with error bars), compared with similar ratios for SN II (green points) and SN Ia; SN II are the result of the evolution of massive young stars found in starburst regions, SN Ia are associated with old stellar populations (from Baldi et al. 2006b).

### 7.3.2 The Hot ISM of Early-Type (Elliptical and S0) Galaxies

Early-type galaxies (ETGs) are characterized by an old stellar population. ETGs are believed to be the end product of galaxy evolution, and to originate from major mergers of gas-rich disk galaxies, with subsequent accretion of smaller mass companions (De Lucia et al. 2006; Oser et al. 2012). The hot ISM and halos of normal elliptical galaxies were discovered thanks to the imaging capabilities of the *Einstein Observatory* (Forman et al. 1985; Trinchieri & Fabbiano 1985; see reviews by Fabbiano 1989, 2012 and references therein). This discovery went against the accepted picture in astronomy at the time, that winds would occur in ETGs, dissipating the gaseous stellar ejecta outside the parent galaxy (e.g., Faber & Gallagher 1976; Mathews & Baker 1971). While many studies with *Einstein* (and the following X-ray observatories *ROSAT*, *ASCA*) ensued, it is only with *Chandra* that these hot gaseous components can be studied in detail, with relatively uncontroversial results (see review, Fabbiano 2012).

With the sub-arcsecond imaging of *Chandra*, and its spectral capabilities, the populations of LMXBs in ETGs can be detected and separated from the hot gaseous emission, both spatially and spectrally (Section 7.2). This capability has led to the resolution of previous hotly debated ambiguities, regarding the amount of hot ISM and halos in different ETGs, and has allowed the physical and chemical characterization of these hot halos. Studies of the interactions between hot ISM, gravity, and nuclear activity, based on these data, are setting stringent constraints on the amount of dark matter and nuclear feedback in ETGs.

Below we will review some of these topics, with emphasis on the observational picture of ETGs and its implications for their evolution. This discussion is complementary to that of Chapters 9 and 10, where a more in-depth discussion of the physics of the hot halos can be found (see also the reviews by Pellegrini 2012; Sarazin 2012; and Ciotti & Ostriker 2012).

*7.3.2.1 Scaling Relations of ETGs*

The first question following the discovery of hot halos in some elliptical galaxies in the Virgo cluster (Forman et al. 1979) was: how widespread are these halos? (Trinchieri & Fabbiano 1985), followed by the obvious corollary: how can we constrain the physical properties and the evolution of these halos, taking into account different scenarios for the heating of the gas (e.g., Forman et al. 1985; Canizares, Fabbiano & Trinchieri 1987)? A basic tool for these studies was the comparison between the integrated X-ray luminosity ($L_X$) of the galaxies and their integrated stellar emission ($L_B$, earlier studies used B-band integrated stellar emission, later studies used the K-band emission, which is more representative of the stellar population of these galaxies).

Since the emission of LMXB populations was also included in the total X-ray emission, it was highly controversial if the pre-*Chandra* $L_X$-$L_B$ diagrams truly represented the behavior of the hot ISM as a function of the stellar mass of the galaxies. Moreover, early measurement of the temperatures of these halos were biased by the mixing in of the hard LMXB spectra, especially in galaxies with relatively smaller amounts of hot halos (those with lower $L_X$ / $L_B$ ratios; Kim et al. 1992). While studies with *ROSAT* and *ASCA* have probed some of these issues, the angular resolution of the *Chandra* telescope was needed to change the observational paradigm (see Fabbiano 2012).

With *Chandra* ACIS, Boroson, Kim and Fabbiano (2011; BKF) derived scaling relations for the different components of the X-ray emission of a sample 30 nearby ETGs, spanning a range of X-ray-to-optical ratios. The K-band near-IR integrated emission was used as a proxy of the total stellar mass of each galaxy. BKF detected samples of LMXBs in these galaxies and subtracted their contributions from the images. They subtracted the contribution of fainter LMXBs, below the detection threshold, following two approaches: (1) extrapolating to lower luminosities the LMXB X-ray luminosity function (see Section 7.2.1); and (2) spectral analysis, since the LMXB spectrum is harder than that of the hot ISM. They also considered the contribution of the coronally active binaries (ABs) and cataclysmic variables (CVs) in these galaxies, using *Chandra* observations of the bulge of M31 and of M32 to model this unresolved emission. Finally, the emission from sources detected at the galaxy nucleus, which could be X-ray faint AGNs, was also subtracted.

The BKF $L_X$-$L_K$ diagram of the hot halo is shown in the left panel of Fig. 7-7 (red points) together with the total ETG $L_X$ (black circles; effectively the pre-*Chandra* scaling relation), and the contributions of LMXBs (blue dots), ABs+CVs (black line) and nuclear sources (green triangles). There is no correlation between the nuclear luminosity (green points) and $L_K$. For the hot gas, the $L_X$

-$L_K$ points (red) follow a much steeper relation than in the pre-*Chandra* studies (the circles), but still have considerable spread. The LMXB contribution is correlated with the stellar mass (represented by $L_K$), although a secondary correlation with $S_N$, the number of GC per unit stellar mass in a galaxy (see Section 7.2), was also found by BKF. The AB+CV contribution is by construction correlated with $L_K$. The average LMXB contribution to the integrated $L_X$ is 10 times larger than that from the unresolved AB+CV emission:

$$L_X(LMXB)/L_K = 7.6 \times 10^{28} \times S_N^{0.334} \; erg \; s^{-1} \; L_K^{-1} \quad \text{and} \quad L_X(AB+CV)/L_K = 9.5 \times 10^{27} \; erg \; s^{-1} \; L_K^{-1}$$

One of the most striking results of the BKF study is the strong positive correlation found between the luminosity and temperature of the hot gas component (right panel of Fig. 7-7), which follows the best-fit relation $L_X$ (gas) ~ $T^{4.6 \pm 0.7}$ (green line in the figure; the cyan line is the similar best-fit relation obtained excluding ETGs with $L_X$ (gas) < $10^{39}$ erg s$^{-1}$, where the errors are larger). The yellow line represents the relation (similar, but shifted towards higher $L_X$) for the central galaxies of groups and clusters. A correlation between gas luminosity and temperature is expected for gravitationally confined hot halos (see also Section 7.3.2.2).

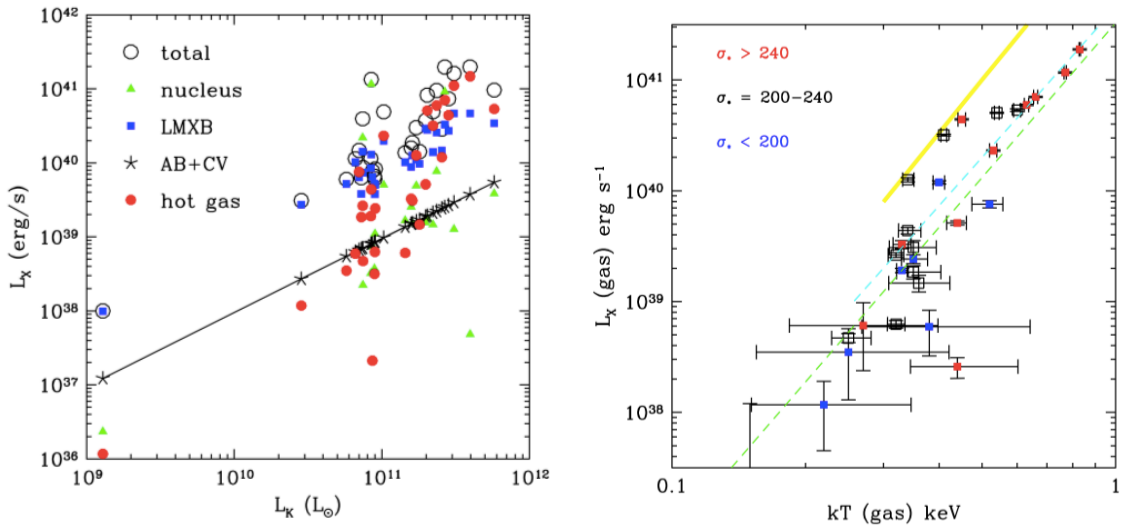

Fig. 7-7– From Boroson, Kim & Fabbiano 2011. Left: $L_X$ -$L_K$ diagram for the sample of 30 nearby ETGs observed with *Chandra*, including the total $L_X$ of each ETG (circles), and the contribution of each component as indicated. Right: $L_X$ – kT diagram for the hot gas. The dashed lines are best-fit relations for the entire sample and galaxies with $L_X$ (gas) > $10^{39}$ erg s$^{-1}$, the yellow line represents the relation of cD galaxies and groups.

The $L_{X,GAS}$-$L_K$ and $L_{X,GAS}$-$T_{GAS}$ relations were further investigated by Kim & Fabbiano (2015) using a larger sample of ETGs, the 61 ATLAS3D E and S0 galaxies observed with *Chandra*, including *ROSAT* results for a few X-ray bright galaxies with extended hot gas. This work uncovered a dependence in these relations, on the structural and dynamical properties of ETGs, suggesting that the correlations are carried by the 'core' ETGs in the sample, ETGs with central surface brightness cores, slow stellar rotations, and uniformly old stellar populations. For these galaxies, the $L_{X,GAS}$ ~ $T_{Gas}^{4.5+/- 0.3}$ correlation extends down into the $L_{X,Gas}$ ~ $10^{38}$ erg s$^{-1}$ range, where simulations predict the gas to be in outflow/wind state (Negri et al. 2014), with resulting $L_X$ values much lower than observed. Instead, the observed correlation may suggest the presence of small bound hot halos even in this low luminosity range.

The $L_{X,GAS} \sim T_{Gas}^{4.5+/-0.3}$ correlation of core ETGs is consistent with the presence of virialized hot halos. The virial theorem requires that if the hot gas is in equilibrium in the gravitational potential, $M_{Total} \sim T_{Gas}^{3/2}$. Kim & Fabbiano (2013; see also Forbes et al. 2017) found a tight relation between the gas X-ray luminosity and the dynamical galaxy mass: $L_{X,Gas}/10^{40}$ erg s$^{-1}$ = $(M_{Total}/3.2 \times 10^{11} M_\odot)^3$. Substituting $L_{X,Gas} \sim T_{Gas}^{4.5}$, we obtain the virial relation. Therefore, in these galaxies the dark matter is primarily responsible for retaining the hot gas.

Among the gas-poor galaxies, which also tend to be cuspy in their internal isophotes, the scatter in the scaling relations increases, suggesting that secondary factors (e.g., rotation, flattening, star formation history, cold gas, environment, etc.) may become important (Kim & Fabbiano 2015). In these galaxies, $L_{X,GAS}$ is < $10^{40}$ erg s$^{-1}$ and is not correlated with $T_{GAS}$. The $L_{X,Gas}$-$T_{Gas}$ distribution is a scatter diagram similar to that reported for the hot interstellar medium (ISM) of spiral galaxies (Li & Wang 2013), suggesting that both the energy input from star formation and the effect of galactic rotation and flattening may disrupt the hot ISM.

Fig. 7-8, from Kim & Fabbiano (2015), compares the observed distributions of ETGs in the $L_{X,Gas}$ – $T_{Gas}$ plane, with those of cD galaxies (the dominant galaxies of groups and clusters, sitting at the bottom of the potential well imposed by the group dark matter, e.g., M87 in the Virgo Cluster), and with groups and clusters of galaxies. The dashed line shows the expectation for gravitational confinement alone ($L_{X,GAS} \sim T_{GAS}^2$), which does not even represent the correlation for galaxy clusters $L_{X,Gas} \sim T_{Gas}^3$ (see Arnaud & Evrard 1999; Maughan et al. 2012). Non-gravitational effects may be responsible for the increasing steeper relations for less massive systems. The $L_{X,Gas}$-$T_{Gas}$ correlation of core elliptical galaxies ("Pure Es" in the figure) is similar to that found in samples of cD galaxies and groups, but shifted down toward relatively lower $L_{X,Gas}$ for a given $T_{Gas}$. Enhanced cooling in cDs, which have higher hot gas densities and lower entropies, could lower $T_{Gas}$ to the range observed in giant Es, a conclusion supported by the presence of extended cold gas in several cDs. In the smaller halos of ETGs the effects of supernova heating and nuclear feedback would be the strongest, depleting the hot halos (i.e. lowering $L_{X,Gas}$).

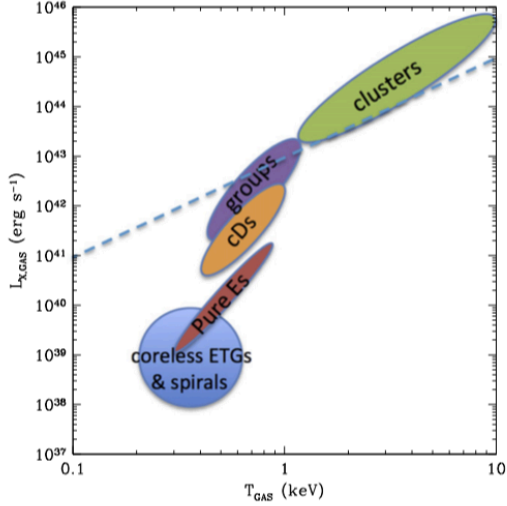

Fig. 7-8. From Kim & Fabbiano (2015). Comparison of the $L_{X,GAS}$ – $T_{GAS}$ of different classes of objects, as indicated. No correlation is found in spiral galaxies and coreless (cuspy) ETGs. Normal core E galaxies follow a tight $L_{X,GAS} \sim T_{GAS}^{4.5}$ correlation. CD galaxies and groups follow a similar correlation, but shifted towards higher $L_{X,GAS}$. Cluster of galaxies follow a flatter $L_{X,GAS} \sim T_{GAS}^{3}$ relation. The dashed line shows the expectation for gravitational confinement alone ($L_{X,GAS} \sim T_{GAS}^{2}$).

*7.3.2.2 Constraints on the Binding Mass of ETGs*

As first applied to M87, the central galaxy of the Virgo cluster, the binding mass of hot halos in gravitational equilibrium can be measured using the equation (Fabricant et al. 1980):

$M(r) = - kT_{gas} / G\mu m_H \, (d \log \rho_{gas} / d \log r + d \log T_{gas} / d \log r) \, r$

Where M(r) is the binding mass within radius r, $T_{gas}$ and $\rho_{gas}$ are the gas temperature and density at the same radius, k is the Boltzmann constant, G is the gravitational constant, μ is the molecular weight and $m_H$ is the mass of the hydrogen atom. Based on this equation, the mass enclosed within the outer detected halo radius is a function of four measurable quantities: temperature, density, and their radial gradients. The uncertainty on these mass measurements derives from the uncertainties on each of these four quantities.

The use of the hot halos of ETGs as a way to measure the gravitational mass of the galaxy has been intensely debated. Issues include the uncertainties in these measurements, the potentially large biases resulting from the contaminations of the halos with undetected LMXB populations, and the physical status of the halos themselves, which may not be solely in gravitational equilibrium. Large halos can be affected by nuclear feedback at small radii and by interaction with their environment at large radii, while small halos could be escaping as galactic winds (see e.g. the case of the Fornax A galaxy, NGC 1316, Kim & Fabbiano 2003; and reviews, Fabbiano 2012; Statler 2012; Buote & Humphrey 2012).

While this discussion continues, recent *Chandra* work has sought to compare dynamical mass measurements (from stellar, planetary nebulae and GC kinematics) with X-ray mass measurements. The purpose of these comparisons is twofold: to provide a way to establish the validity of X-ray mass measurements, and to constrain the physical state of the halos where departures of the X-ray mass estimate from the dynamical mass estimate are observed. For this work, the halos observed with *Chandra* are 'cleaned' of LMXB contaminants (see Section 7.3.2.1). In some cases, lower angular resolution *XMM-Newton* data was also used at large radii, where the effect of the LMXB population is small.

In NGC 4649, a giant E galaxy in Virgo with an extensive hot halo, Paggi et al. (2014, 2017b) find a clear difference in the X-ray and dynamical mass profiles that can be related to the effect of nuclear feedback from the expanding radio lobes in the hot halo (Fig. 7-9). This non-thermal pressure amounts to ~30% of the gravitational pressure, counter-balancing somewhat the effect of gravity in the inner radii and resulting in an apparently smaller X-ray mass than dynamical mass.

In NGC 5846, Paggi et al. (2017) find significant azimuthal asymmetries in the X-ray mass profiles. Comparison with optical mass profiles suggests significant departures from hydrostatic equilibrium, consistent with bulk gas compression and decompression due to sloshing on ~15 kpc scales. Only in the NW direction, where the emission is smooth and extended, do they find consistent X-ray and optical mass profiles, suggesting that the hot halo is not affected by strong non-gravitational forces. These authors also note how the results are dependent on the assumptions made on the metal abundance of the gas, because abundance and temperature are coupled in the spectral fit.

The above examples demonstrate how careful an analysis is needed when attempting to characterize the properties of the hot gas, and measure the galaxy mass, in each individual case. Instead, as discussed in Section 7.3.2.1, the scaling relations of ETGs, in comparison with dynamical mass measurements, provide a more encouraging picture, at least for a class of ETGs: the core elliptical galaxies, where the scaling relations appear strongest (Kim & Fabbiano 2013, 2015, Forbes et al. 2017).

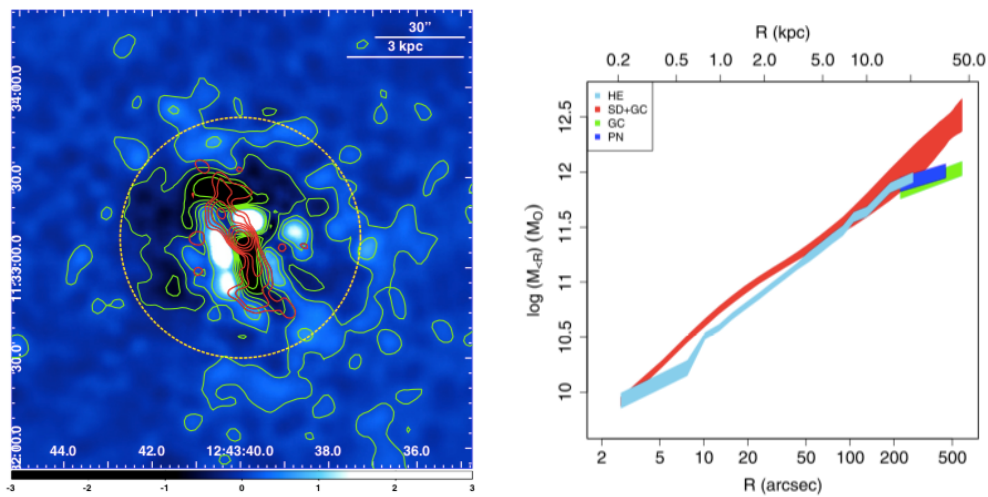

Fig 7-9 – From Paggi et al. (2014). Left: Ridges in the hot gas distribution of NGC 4649, resulting from the interaction with the nuclear radio source (the red contours). Right: Comparison of the radial mass profile from the hot gas in the assumption of hydrostatic equilibrium (pale blue) with mass profiles derived from the kinematics of star, planetary nebulae and globular clusters. Note the smaller values of the X-ray derived mass in the central ~3 kpc, where the gas is subject to the pressure of the expanding radio jet/lobes (see left panel).

*7.3.2.3 Metal Abundances of the Hot Halos*

The physical evolution of the hot halos is closely linked to their chemical evolution, since these halos are enriched in metals by stellar and supernova ejecta. In particular, large hot halos in gravitational equilibrium should have Fe content commensurate to the integrated output of SNe Ia over their lifetimes. In these halos, the Fe to alpha element ratios should be solar or higher, unless inflow of intra-cluster gas, enriched by SNe II-powered winds early in the galaxy's lifetime, alter these values (David et al. 1991; Ciotti et al. 1991; Renzini et al. 1993; Arimoto et al. 1997). The X-ray spectra should contain emission lines revealing the imprint of these metals.

As discussed in Section 7.3.1.1 for the metal abundances of the hot ISM and outflows of star-forming galaxies, most of the early measurements in ETGs returned puzzling sub-solar abundances. More recent work, both based on a careful analysis of *ASCA* spectra (Matsushita et al. 2000), and on *Chandra* and *XMM-Newton* observations, has produced results more in keeping with the expectations (Kim & Fabbiano 2003; 2004). The problem with past results resided both in the contamination of the spectra by the LMXB population, and by radial temperature and abundance gradients in the hot halos. In NGC 507, in particular (Fig. 7-10; Kim & Fabbiano 2004), the super-solar Fe abundances at small radii are consistent with chemical evolution models, of enrichment by SNIa throughout the lifetime of the galaxy, after the initial star formation when SNII are responsible for the enrichment.

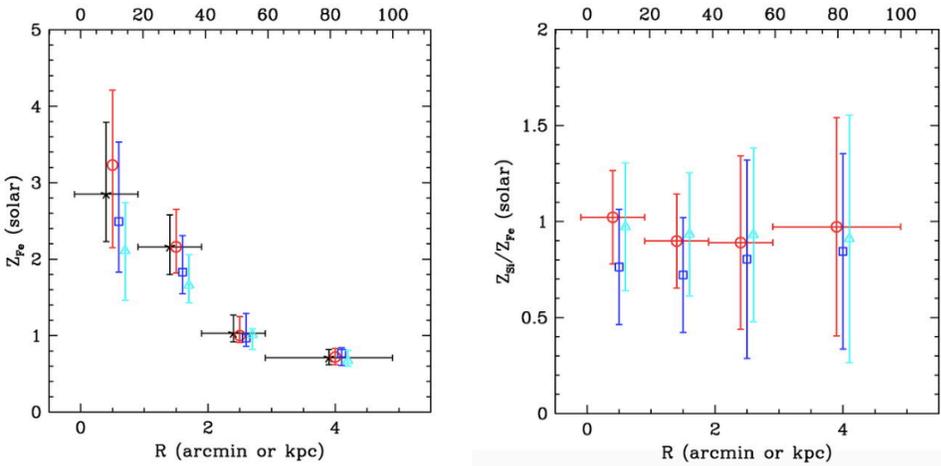

Fig. 7-10 – From Kim & Fabbiano (2004). Radial distributions of (a) Fe abundance and (b) Si-to-Fe abundance ratios measured with *Chandra* data for NGC 507. Different symbols indicate different ways of tying the elements in the spectral analysis.

The ratio of Si to Fe is also consistent with this picture, because significant larger ratios derive from SNII yields. For an in-depth discussion of observational and analysis issues, see Kim (2012). A discussion of the chemical evolution of the hot gas of ETG, enriched by stellar winds and supernovae throughout the lifetime of the galaxy, can be found in Pipino (2012).

*7.3.2.4 ETGs at Higher Redshift*

The studies of ETGs discussed so far were all based on nearby galaxies, mostly within 30 Mpc. To study the evolution of these systems, it is necessary to look at the higher redshifts explored with *Chandra* deep and medium-depth surveys. Studies including both ETGs detected in the *Chandra* Deep Fields, and stacking analysis of undetected galaxies in these fields, suggest no or mild evolution of the X-ray luminosity (relative to the optical luminosity) up to z~1.2 (Tzanavaris & Georgantopoulos 2008; Lehmer et al. 2007; Danielson et al. 2012). The mild increase in luminosity with redshift for ETGs was confirmed by the stacking analysis of a sample of K-band selected galaxies in the COSMOS field surveyed with *Chandra* by Jones et al. (2014). These authors suggested mechanical heating from radio AGNs as the heating mechanism.

Civano et al. (2014) took a more direct approach, comparing the sample of ETGs detected with *Chandra* in the C-COSMOS field (Elvis et al. 2009) with the local universe ETGs studied with *Chandra*. The C-COSMOS detections include 69 ETGs with $L_X$ ~$10^{40}$ - $10^{43.5}$ erg s$^{-1}$, and redshift z≤1.5. The optical spectra are consistent with a passive old stellar population, but a few 'possible' AGNs (Active Galactic Nuclei, see Chapter 8) are included. As we know from detailed studies of nearby galaxies, the X-ray emission of ETGs is complex: it includes the integrated emission of the stellar LMXB population, the emission of hot gaseous ISM and halos, and possible AGN emission from the nuclear super-massive black hole (see Sections 7.2 and 7.3.2.1). The purpose of Civano et al. was to explore the z-evolution of hot halos and the occurrence of low-luminosity AGNs. To this end, the expected LMXB contribution was subtracted using the BFK scaling relation (Section 7.3.2.1), derived from the ETGs in the local universe, modified to take into account the expected z-evolution of these populations (Fragos et al. 2013a). The resulting counts were converted to luminosity assuming a typical range of hot halo temperatures, and compared with the BKF $L_X$(gas) – $L_K$ diagram of local ETGs (Fig. 7-11).

Civano et al. find that most galaxies with estimated $L_X$ < $10^{42}$ erg s$^{-1}$ and z < 0.55 follow the $L_{X, gas}$-$L_K$ relation of local universe ETGs (the red Local Strip in Fig. 7-11). The stacking experiment of Paggi et al. (2016; see Section 7.4.1) shows that the average X-ray spectra of Local Strip galaxies are soft, consistent with gaseous emission. All the X-ray ETGs with stellar age > 5 Gyr follow reasonably well the Local Strip, suggesting that the hot halos are similar to those observed in the local universe (Fig 7-12, right). This result is consistent with the predictions of evolutionary gas-dynamical models including stellar mass losses, supernova heating, and AGN feedback (Pellegrini 2012; Pellegrini et al. 2012). For these galaxies, total masses may be derived using the virial relation of the local sample (Kim & Fabbiano 2013, see Section 7.3.2.1).

There are a few galaxies to the left of this strip (in the green 'X-ray excess' locus in Fig. 7-11). These have X-ray emission well in excess of the $L_{Gas}$ expected from the local sample ETGs given their stellar mass (or K-band luminosity). This excess X-ray emission suggests the presence of a different type of source, neither LMXBs nor hot gaseous halos. Likely candidates are low-luminosity AGNs (see Section 7.4.1). The more luminous ($10^{42}$ erg s$^{-1}$ <$L_X$< $10^{43.5}$ erg s$^{-1}$) and distant galaxies (higher z) present significantly larger scatter (blue strip in Fig. 7-11; see Fig. 7-12 left). These galaxies typically have younger stellar ages. They also tend to have $L_X > 10^{42}$ erg s$^{-1}$ and be over-luminous in X-rays for their $L_K$ when compared to the local sample and older stellar age galaxies (Fig. 7-12). The *Hubble Space Telescope* (*HST*) images of several of these galaxies show close companions, suggesting that galaxy merging may be responsible for the high X-ray luminosity. Merging may enhance the star formation rate (consistent with the younger stellar ages of these ETGs) thus producing a population of luminous X-ray binaries (see Section 7.2.1), enhance the X-ray luminosity of the halo (Section 7.3.1), and also induce nuclear accretion awakening an AGN (Cox et al. 2006; see Section 7.4.2).

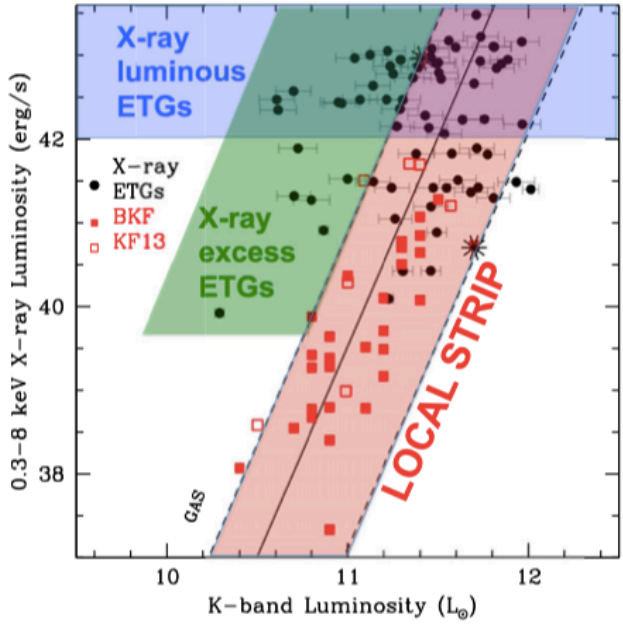

Fig. 7-11 - From Civano et al. (2014). Non-stellar X-ray luminosity (see text) versus K-band luminosity (a proxy of the total stellar mass of the ETG) for the COSMOS detected ETGs (black dots), compared with the local sample gaseous halos (BKF: Boroson et al. 2011; KF13: Kim & Fabbiano 2013; red points). See text for the colored strips.

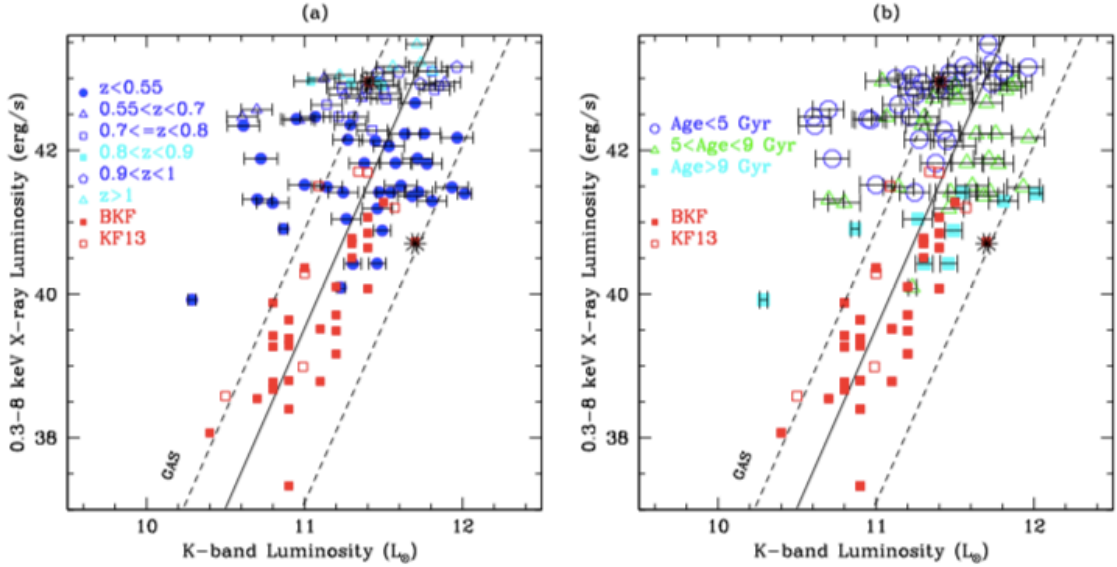

Fig. 7-12 – From Civano et al. (2014). Same as Fig. 7-11, except that the COSMOS ETGs are labelled according to their redshift (left panel) and average stellar age (right panel). The dashed lines identify the boundaries of the Local Strip of Fig. 7-11.

### 7.4 Nuclear BHs and AGNs

Supermassive black hole (SMBHs), with masses $10^8$ M$_\odot$ or larger, are ubiquitous in giant galaxies, and their mass is correlated with the stellar mass of the galaxy bulge (Magorrian et al. 1998; Gültekin et al. 2009). This relation suggests a 'co-evolution' scenario, with the growth of the SMBH linked to the growth of the host galaxy. In the picture of growth via merger and accretion, simulations show that the SMBH of one or both merging galaxies undergoes periods of accretion and activity, becoming an 'active galactic nucleus' – AGN (Hopkins et al. 2008; Volonteri et al. 2016). The AGN, in turn, interacts with the host galaxy, either stimulating or dampening star formation (see Heckman & Kauffmann 2011 and references therein).

Below, we discuss briefly two aspects of nuclear properties that are relevant for having a complete picture of normal galaxies: (1) the presence of low-luminosity 'hidden' AGNs in normal galaxies; and (2) the link between galaxy merging and SMHB activity and growth. This discussion is complementary to Chapter 8, which addresses the X-ray properties of AGNs.

### 7.4.1 Hidden AGNs in normal galaxies

Two still open questions are: (1) what is the full range of nuclear BH masses, i.e., how small can a nuclear black hole be? Do dwarf galaxies host smaller mass nuclear BHs? And (2) what is the extent of nuclear activity in the universe, i.e. the cycle of accretion and starvation of the BH? Answering the first question is relevant for constraining theories of the formation and characteristics of SMBH seeds in the early universe (Volonteri 2012). As for the second question, since normal galaxies host nuclear massive black holes (Magorrian et al. 1998), the distinction between normal galaxies and AGNs is a matter of how much the nuclear BH is accreting.

Observationally, AGNs have only been considered those sources detected in X-ray surveys with $L_X > 10^{42}$ erg s$^{-1}$, since below this threshold the X-ray luminosity can be explained with stellar and hot ISM emission (Fabbiano 1989). However, with *Chandra* one can explore more fully the range of nuclear emission. In nearby galaxies that can be resolved in their emission components, low-luminosity nuclear sources are frequently detected, associated with SMBH masses ranging from the 4.1 +/- 0.6 × 10$^6$ M$_\odot$ SMBH at the center of the Milky Way (Ghez et al. 2008) to ~ 10$^{9-10}$ M$_\odot$ SMBHs in giant ETGs (e.g., Paggi et al. 2014). For more distant galaxies, even if the emission components cannot be resolved via imaging, *Chandra* studies of the properties and scaling laws of the XRB populations and hot halos (see Section 7.2 and 7.3) provide the tools for estimating the AGN contribution to the X-ray luminosity. Low-luminosity nuclei may be too faint or too obscured to leave an AGN signature in the optical spectra, and therefore may be missed by multi-wavelength survey, such as COSMOS, but can be revealed by X-ray studies.

Paggi et al. (2016; see Fig. 7-13) performed a stacking experiment on the 6388 ETGs undetected by *Chandra* in the C-COSMOS sample, by adding up the *Chandra* data of galaxies in bins of optical luminosity ($L_K$) and redshift (z). The expected LMXB contribution to the stack was subtracted, following the procedure of Civano et al. (2014; see Section 7.3.2.4). Similar stacking experiments, using galaxies in the COSMOS survey, were performed by Mezcua et al. (2016) for the ~50,000 undetected star-forming dwarf galaxies with redshift up to z = 1.5, and Fornasini et al. (2018) for 75,000 giant star-forming galaxies between redshift 0.1 and 5.

In dwarf ETGs [log ($L_K$) < 10.5 in solar units], Paggi et al. (2016) find significant emission above that expected from the LMXB populations. Moreover, this emission has hard average X-ray spectra, consistent with AGN emission. The luminosities (a few 10$^{39}$ – 10$^{41}$ erg s$^{-1}$) are consistent with those produced by inefficient accretion (10$^{-5}$ - 10$^{-4}$ of the Eddington rate) onto black holes of mass ~ 10$^6$ - 10$^8$ M$_\odot$. In star-forming dwarfs, with stellar mass < 10$^{9.5}$ M$_\odot$ Mezcua et al. (2016) report an X-ray excess that suggests nuclear accreting BHs, with average nuclear X-ray luminosities in the range 10$^{39}$-10$^{40}$ erg s$^{-1}$ and inferred masses ~1- 9 × 10$^5$ M$_\odot$. These nuclear BH masses derived from the stacking of survey data are in the range of the intermediate mass BH AGNs studied by Greene & Ho (2007), and of those inferred by *HST-Chandra* studies of nearby single dwarf galaxies (e.g., Baldassarre et al.2017), and of low-luminosity AGNs in small-bulge galaxies (Chilingarian et al. 2018).

Highly obscured hidden AGNs in giant galaxies are reported by Paggi et al. (2016) in the high redshift ETG COSMOS sample (Fig. 7-13), in particular those with X-ray luminosity > 10$^{42}$ erg s$^{-1}$. Similarly, hidden highly obscured AGNs at z>1.5 are also found in the stacking analysis of star-forming giant galaxies by Fornasini et al. (2018). In both cases these AGNs are not visible in the COSMOS optical-IR colors and spectra.

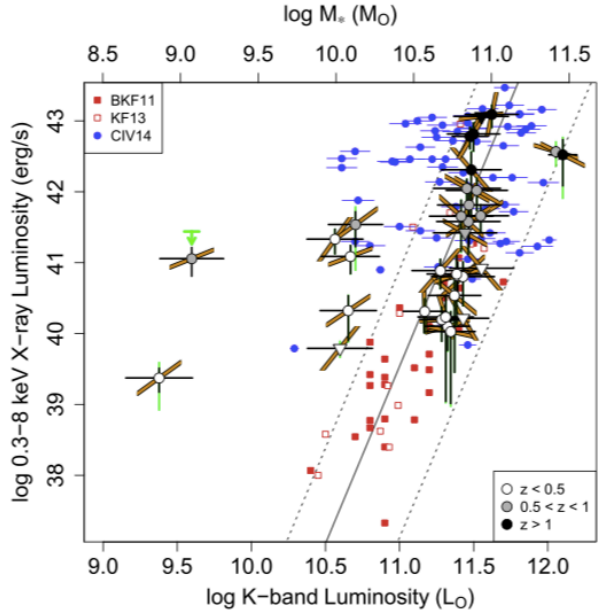

Fig. 7-13 – From Paggi et al. (2016). Average non-stellar (LMXB contribution subtracted) X-ray luminosity of $L_K$ stacks (circles, shaded according to the average redshift of the stack), for non-X-ray-detected COSMOS ETG. The blue points are the COSMOS detections (Civano et al. 2014, Section 7.3.2.4), and the red points are the near universe gaseous halos. The yellow diagonal bars across the points represent the average power-law spectrum of the stack. Hard spectra (up-pointing power-laws) are found both in the highest stellar mass ($L_K$) highest $L_X$ stacks, and in the X-ray excess dwarf ETG stacks. The stacks that follow the local strip of Fig. 7-11 (between dashed lines) tend to have soft X-ray spectra (down-pointing power-law), consistent with the emission of hot halos.

### 7.4.2 AGNs in Merging Galaxies

Simulations (Hopkins et al. 2008) suggest that galaxy merging and the resulting AGN activity are key steps of galaxy evolution. During this process, the SMBH may be "buried" by thick molecular gas, which feeds the SMBH at high rates, causing the birth of an obscured Compton Thick (CT; Risaliti et al. 1999) AGN. Uncontroversial statistics are missing on the occurrence of dual AGNs that may then merge into a single SMBH after releasing gravitational waves (Shannon et al. 2013). However, some evidence of merger-related AGN activity has been provided by *Chandra* observations.

The first dual CT AGNs were discovered with *Chandra* in the merger NGC 6240 (Komossa et al. 2003). Both nuclei are detected in hard continuum and have the large 6.4 keV Fe Kα emission lines characteristic of CT AGNs. Extended, hard X-ray continuum, and Fe XXV line emission in this galaxy has be related either to shocks caused by the intense star-formation activity near the nuclei or to AGN winds (Wang et al. 2014).

Deep *Chandra* observations of the merger Arp 220 (Paggi et al. 2017a) clearly show that the nuclear regions are dominated by Fe XXV emission in X-rays, suggesting either highly shrouded nuclei or just the result of intense star formation. One active AGN contributes to the X-ray luminosity of the merger Arp 299 (Anastasopoulou et al. 2016). Ultraluminous infrared galaxies (ULIRG) could be the result of recent mergers. A *Chandra* survey of these galaxies found widespread Fe XXV emission (Iwasawa et al. 2009).

While most ULXs have been convincingly related to the X-ray binary population of the host galaxy, some very luminous ones may indicate a merging remnant. A compelling example of an accreting black hole with a mass of $\geq 10^4$ M$_\odot$, which may be the stripped remnant of a merging dwarf galaxy, is the very luminous off-nuclear X-ray source discovered in the Seyfert galaxy NGC 5252 (Kim et al. 2015). With an X-ray luminosity of $1.5 \times 10^{40}$ erg s$^{-1}$, this source is associated with radio emission and optical emission lines (at the redshift of NGC 5252). The flux of [O III] appears to be correlated with both X-ray and radio luminosity in the same manner as ordinary AGNs.

### 7.5 AGN-galaxy interaction in nearby spiral galaxies

As discussed in Section 7.4, nuclear SMBHs are ubiquitous in galaxies. When accreting, they give rise to AGNs, which produce a significant fraction of (or exceed) the total energy output of the galaxy from other sources. The physical processes related to AGN activity are discussed in Chapter 8. However, both AGN radiation and the mechanical energy associated with jets and outflows have an impact on the host galaxy. AGN feedback is believed to be one of the important factors in galaxy evolution (see Section 7.4). Deep *Chandra* observations of nearby galaxies provide a new perspective on this interaction. Nuclear radio sources (jet and lobes) in elliptical galaxies carve cavities in the hot ISM (see Section 7.3.2.2 and Fig. 7-9), and effectively stop cooling flows. These interactions are extensively discussed in Chapters 9 and 10. Here, instead, we address the interaction of the AGN with the colder ISM of spiral galaxies.

The X-ray emission of AGNs in the ~0.3-8.0 keV *Chandra* energy range is characterized by a spatially extended soft component at energies < 2.5 keV, which is dominated by line emission, and a harder continuum component at higher energies. In some heavily obscured AGNs, prominent Fe Kα 6.4 keV neutral lines are also seen. In some cases, Fe XXV at 6.7 keV has also been detected. Fig. 7-14 shows the *Chandra* X-ray spectrum of one such AGN, observed with *Chandra* ACIS (Fabbiano et al. 2017): ESO 428-014, which has a heavily absorbed ($N_H > 10^{25}$ cm$^{-2}$) Compton Thick (CT) Seyfert Type 2 nucleus (Risaliti et al. 1999). In the 'standard model' (e.g., Urry & Padovani 1995), the soft component would by emitted by clouds in the galaxy ISM, photoionized by the AGN (Bianchi et al. 2006), while the hard continuum and the Fe Kα lines are believed to arise from the interaction of hard photons from near the nuclear SMBH with a small-scale ~1-100 pc circumnuclear obscuring torus (e.g., Matt et al. 1996; Ghandi et al. 2015).

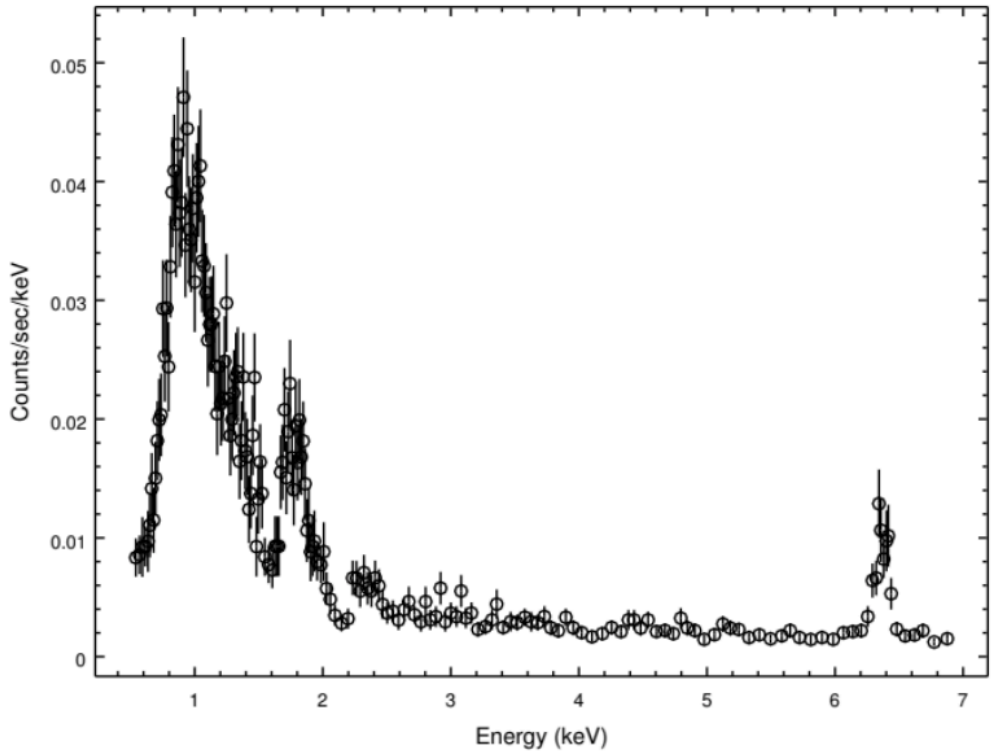

Fig. 7-14– *Chandra* ACIS observed spectrum of ESO 428-G014 (Fabbiano et al. 2018a). It includes all the photons detected within a circle of 8" radius centered on the peak of the hard (>3 keV) nuclear source.

Below we discuss how the 'standard model' paradigm is being explored, and in some cases challenged, by *Chandra* images of the different spectral components of AGNs. This work makes use of the ultimate (~1/4 arcsec) resolution of the *Chandra* telescope, and of multi-wavelength comparison with similar angular resolution radio (*VLA*), optical (*HST*), and millimeter (*ALMA*) images. We first discuss the related observational issues and methods (Section 7.5.1). We then summarize what these observations imply for the nature of the soft diffuse emission (energies < 2.5 keV; Section 7.5.2), and of the hard continuum and Fe K$\alpha$ emission (Section 7.5.3). Finally, we discuss the resulting constraints on the nature of the obscuring torus of the standard model (Sections 7.5.4 and 7.5.5).

### 7.5.1 Methods

Observationally, the sub-arcsecond angular resolution of the *Chandra* telescope allows the resolution of physical regions on a scale of of ~50-100 pc in galaxies at distances of ~10-20 Mpc. However, there are some obstacles to overcome, both of them linked to the instrument of choice for most of these studies, the ACIS. This CCD detector can measure both energy and position of the incoming photons (as well as the time of arrival, which is not relevant for the study of extended features). However, the CCD pixel is a 0.49" side square, and therefore under-samples the *Chandra* PSF. Moreover, the relatively slow CCD readout time (3.2 s if reading out the full chip) results in 'pileup' of multiple photons in a single pixel from an intense point source (such as a bright AGN), causing both spectral and spatial unwelcome effects [see the Chandra Proposers' Observatory Guide (POG)[2]]. The other *Chandra* imager, the HRC (see POG), has instrumental pixels smaller than the mirror PSF, but a worse overall flux-to-counts response than the ACIS (so that much longer exposure times are needed), and virtually no energy resolution.

These problems can be overcome with:

(1) A careful selection of the AGNs suited for these studies (e.g., Fabbiano et al. 2018a). To minimize the effects of ACIS pileup, obscured and highly obscured AGN are the best, so that the nuclear photons would not be directly visible at the low energies (<2 keV). At these low energies, the extended emission is prevalent (Section 7.5.2). The obscuring clouds form a natural coronagraph.

(2) Complementing ACIS observations of less obscured AGNs with HRC observations, to image the innermost circumnuclear regions. This method was used by Wang et al.(2009) in their study of NGC 4151, and Wang et al. (2012) in their study of the innermost regions of NGC 1068. Both AGNs cannot be studied with ACIS in the innermost regions, because of pileup.

(3) Using sub-pixel binning to produce images from ACIS data. This method is made possible by the fact that the *Chandra* telescope scans over the source position in a well-known Lissajous figure pattern (see POG). The method is the continuous limit of the 'multi-drizzle' technique employed on *HST*. It was applied to the ACIS data of NGC 4151 by Wang et al. (2011a), who demonstrated that it recovered the spatial information of the HRC observations of the same region (Wang et al. 2009). To illustrate the advantage of subpixel binning, Fig. 7-15, compares the ACIS native image of the nuclear region of ESO 428-G014 (top left panel), with that obtained by binning the same photons in bins 1/16 of the instrument pixel (top right panel).

(4) Image processing with adaptive smoothing and or image reconstruction (e.g. EMC2, Esch et al. 2004) can then be used to enhance visually detected features. The bottom panels of Fig. 7-15 show the 1/16 bin data processed with adaptive smoothing (left) and EMC2 image restoration plus 2 pixel Gaussian smoothing (right). The contours are from the *HST* Hα image (Falke et al. 1996). The correspondence between Hα features and sub-pixel images is remarkable.

---

[2] http://cxc.cfa.harvard.edu/proposer/POG/

(5) There are some issues with the Chandra PSF one has to be careful about. The first is the well-documented 'hook' feature that requires some checking of the images to exclude spurious results[3]. The second is a marked splitting of the image at the core of the PSF at energies >7 keV (Fabbiano et al. 2019).

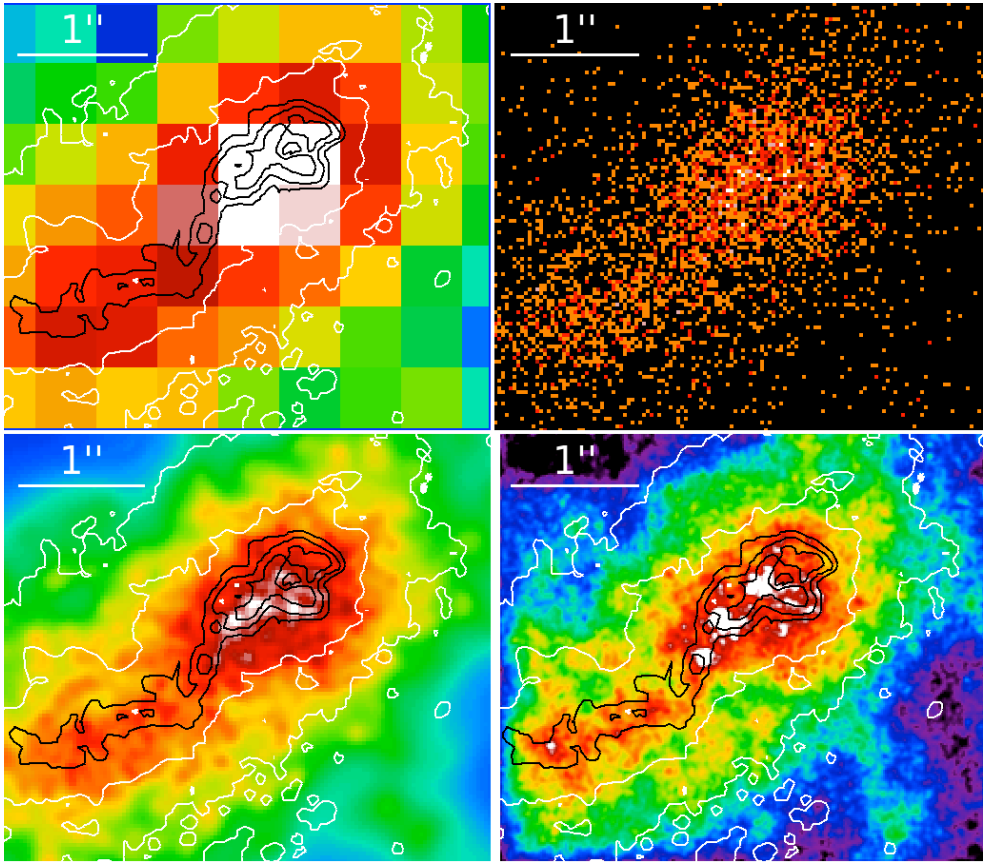

Fig. 7-15 – Data from Fabbiano et al. (2018b). (Top left) Raw 0.3–3 keV ACIS image of the circumnuclear region of ESO 428+G014 (1''=170 pc); (Top right) ACIS image of the same region with subpixel binning (1/16 native pixel); (Bottom left) 1/16 bin image processed with adaptive smoothing; (Bottom right) EMC2 image restoration plus 2 pixel Gaussian smoothing. Hα contours from Falke et al. (1996), superimposed.

### 7.5.2 The Soft (E<2.5 keV) X-ray emission of AGN photoionization cones and soft X-ray constraints on AGN feedback

The soft X-ray emission (energies < ~2.5 keV, see Fig. 7-14) is typically extended and has been associated with the emission of galaxy ISM clouds photoionized by the nuclear photons at a distance from the AGN, or with a wind. This is the same region that also gives rise to the optical narrow-line emission of the ionization cones (e.g., Bianchi et al. 2006; Levenson et al. 2006). Fig. 7-

---

[3] http://cxc.harvard.edu/ciao/caveats/psf_artifact.html

16 shows the striking ionization cone of Mkn 573 (Paggi et al. 2012), as seen with *Chandra* (blue), compared with the [OIII] (*Hubble*, orange), and radio 6 cm (*VLA*, magenta) images.

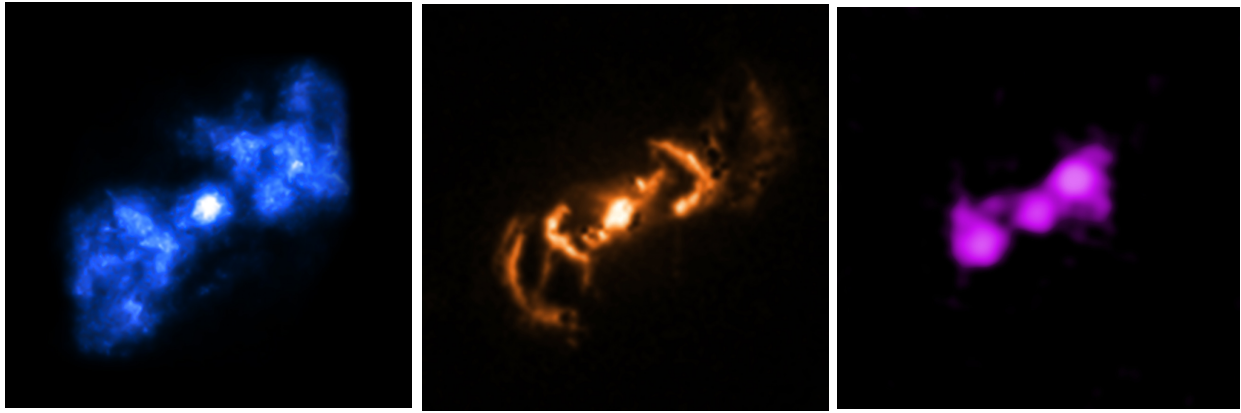

Fig. 7-16 – Left, *Chandra* image of the ionization cone of Mkn 573; Center, [OIII]; Right, radio, 6 cm. Each image is 10'' across, corresponding to ~3.3 kpc at the distance of this galaxy (Paggi et al. 2012).

The ACIS spectra of the 'ionization cones' are generally consistent with photoionization, except for the regions where the radio lobes terminate. Here additional thermal components are required, suggesting shocks driven by the impact of the radio jet on the ISM clouds (see Fig. 7-17, right).

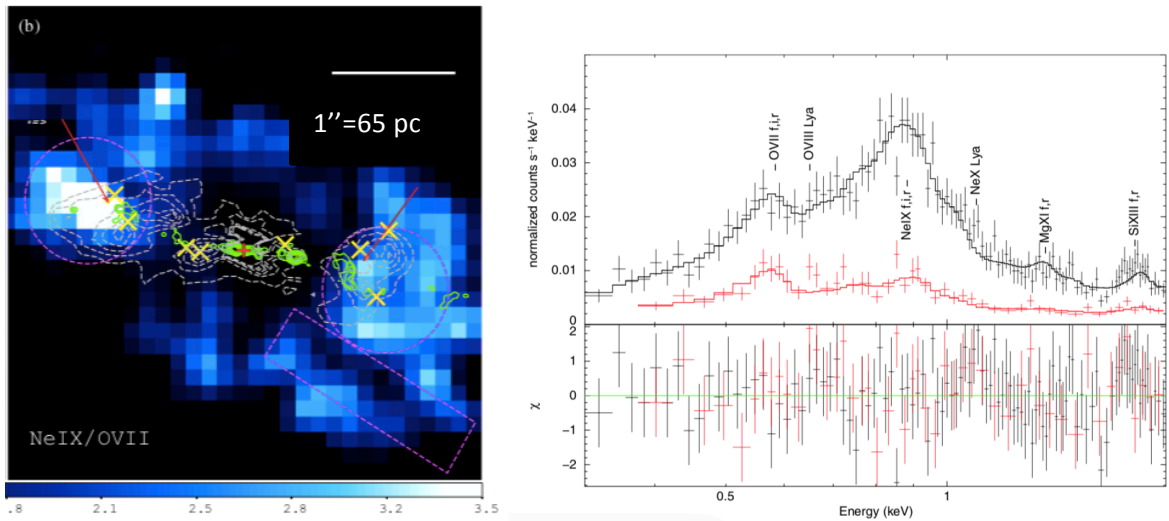

Fig. 7-17 –Left, Ne IX /OVII ratio map from the ACIS observations of the nucleus of NGC4151 (1/8 subpixel binning); green contours represent the radio jet, and dashed contour near-IR [Fe II] emission; yellow crosses mark the locations of the HST clouds with high velocity dispersion. The magenta circles are the extraction regions for the spectrum shown in black in the right panel, while the spectrum from the rectangular area is shown in magenta. The latter can be fitted with photoionization models, while the spectrum from the Ne IX excess areas require an additional thermal component; the best-fit residuals are shown in the bottom of the right panel (Wang et al. 2011b and references therein).

This feature was first observed by Wang et al. (2011b) in the *Chandra* NeIX/OVII line ratio map of NGC 4151, where it appears as localized excesses of the line ratio at a distance ~1'' (65 pc) from the nucleus (Fig. 7-17, left). A similar effect was also observed in Mkn 573 (Paggi et al. 2012).

Measuring individual regions of the ionization cone in both *HST* and *Chandra* images, Wang et al. (2011c) finds that the ratios of [O III]/soft X-ray flux are approximately constant from ~30 pc from the nucleus to the 1.5 kpc radius spanned by their measurements. These ratios are consistent with the average values measured from the soft components of several Seyfert galaxies with *XMM/Newton* (Bianchi et al. 2006). If the [O III] and X-ray emissions both arise from a single photoionized medium, this constant ratio implies a constant ionization parameter (flux/density ratio) and so an ISM density decreasing as $r^{-2}$. This naturally suggests an outflow with a wind-like density profile. The regions corresponding to the termination of the radio jets/lobes have smaller [OIII]/soft X-ray ratios, consistent with the presence of the additional thermal X-ray components that we have discussed above (see Fig. 7-17).

Using spatially resolved X-ray features, Wang et al. (2011c) estimate that the mass outflow rate in NGC 4151 is ~2 $M_\odot$ yr$^{-1}$ at 130 pc from the nucleus and that the kinematic power of the ionized outflow is $1.7 \times 10^{41}$ erg s$^{-1}$, approximately 0.3% of the bolometric luminosity of the active nucleus in NGC 4151. The contribution from the radio-termination shocks is similarly small, suggesting that 0.1% of the jet power is deposited in the ISM (Wang et al. 2011b). These values are significantly lower than the expected efficiency in the majority of quasar feedback models, but comparable to the two-stage model described in Hopkins & Elvis (2010).

In NGC 4151, the AGN-galaxy interaction extends on spatial scales larger than the ionization cones. Wang et al. (2010) discovered soft diffuse X-ray emission in NGC 4151, with a luminosity $L_{0.5-2\ keV} \sim 10^{39}$ erg s$^{-1}$, extending out to ~2 kpc from the active nucleus and filling in a cavity surrounded by cold H I material (Fig. 7-18).

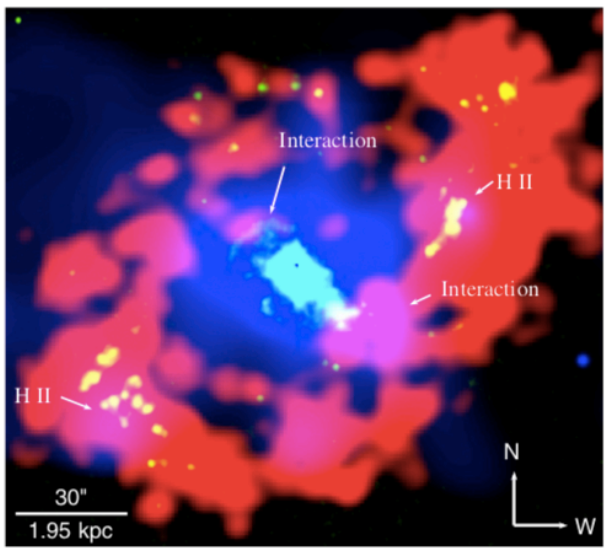

Fig. 7-18 – The hot circumnuclear 2 kpc radius X-ray bubble of NGC 4151 (blue), within the ring of HII emission, in red (Wang et al. 2010).

The best fit to the X-ray spectrum of the X-ray emission of the 'cavity' of NGC 4151 requires either a kT ~ 0.25 keV thermal plasma, or a photoionized component from an Eddington limit nuclear outburst. For both scenarios, the AGN-host interaction must have occurred relatively recently (some $10^4$ yr ago). This very short timescale to the last episode of high activity phase may imply such outbursts occupy ~1% of AGN lifetime

### 7.5.3 A *Chandra* surprise: the extended hard and Fe Kα emission of AGN

As mentioned at the beginning of Section 7.5, it is commonly assumed that the hard continuum emission (>2.5 keV, see Fig. 7-14) and the Fe Kα 6.4 keV line of CT AGNs both originate from the interaction of hard nuclear photons with the dense Compton-thick clouds that constitute a 0.1-~100 pc radius 'obscuring torus' surrounding the nucleus.

Recent *Chandra* observations are beginning to challenge this paradigm, showing that the hard continuum and Fe Kα emission are not confined to a circumnuclear AGN torus. In particular, in the Circinus Galaxy (Arevalo et al. 2014) and in NGC 1068 (Bauer et al. 2015), both of these emissions also have extended components in the direction of the ionization cone (~600 pc in Circinus, >140 pc in NGC 1068). Kiloparsec-scale hard continuum and Fe Kα emission, extended along the ionization cone, have been discovered in the CT AGN ESO 428-G014 (Fabbiano et al. 2017; Fig. 7-19). These results suggest that hard photons from the AGN, escaping along the same torus axis as the soft photons that give rise to the ionization cone, are then scattered / reflected by dense molecular clouds in the galactic disk of ESO 428-G014.

At energies >3 keV, the extended emission in the central 1''.5 (170 pc) radius circumnuclear region amounts to ~30-70% of the contribution of a point source in that area (or ~25-40% of the total counts in the region). Within a 5" radius, the contribution from the extended emission is greater than that from a nuclear point source in the 3-4 keV band. The size of the extended region of ESO 428-G014 decreases with increasing energy in the *Chandra* range, suggesting that the optically-thick scattering clouds are preferentially located at smaller galactocentric radii (Fabbiano et al. 2018a, b), as is the case for the Milky Way molecular clouds (e.g., Nakanishi & Sofue 2006).

The presence of extended hard continuum and Fe Kα emission does not in itself challenge the 'standard model', since this model accounts for AGN photons escaping along the axis of the torus. However, it illustrates how different properties of the ISM of the host galaxies may cause different phenomena. In particular, AGNs in galaxies with thick molecular disks should be more prone to present extended hard components, because they have the molecular clouds needed to interact with the hard photons.

Extended hard components, however, may adversely bias the torus modeling of spectra from X-ray telescopes with inferior angular resolution than *Chandra*, such as *NuSTAR* and *XMM-Newton,* because these spectra will include a substantial fraction of reprocessed photons not belonging to the torus. In ESO 428-G014 the Fe Kα emission within a circle of 8" (~900 pc) radius centered on the nuclear peak is commensurable with that of a nuclear point source centered on that peak (Fabbiano et al. 2018a, b; Fabbiano et al. 2019). High-resolution X-ray images are needed to complement these spectral studies.

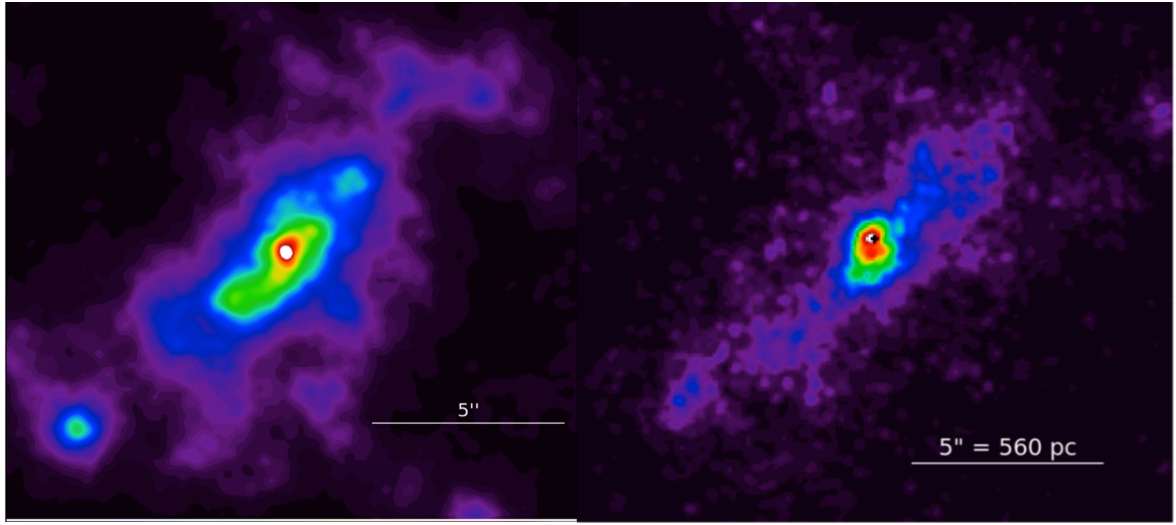

Fig. 7-19. Left, 3-6 keV continuum image of ESO 428-G014 (Fabbiano et al. 2017). Right, Fe Kα image (Fabbiano et al. 2018a). The color intensity scale goes from purple (lowest intensity) to white (highest intensity).

**7.5.4 Imaging the Obscuring Torus**

High resolution *Chandra* imaging of the inner circumnuclear regions of CT AGNs, in the spectral bands representative of the hard continuum (2-6 keV) and Fe Kα (6-7 keV) emission (see Fig. 7-14), has provided some direct observational constraints on the central obscuring structures.

The *Chandra* ACIS observations of NGC 4945 (Marinucci et al. 2012) provided the first opportunity to image circumnuclear regions commensurate with the scale of the obscuring torus. NGC 4945 hosts a highly obscured CT AGN ($N_H \sim 4 \times 10^{24}$ cm$^{-2}$) and is nearby. At the distance of this galaxy (~3.7 Mpc), 1" corresponds to 18 pc. In addition to the ~kpc extended soft emission component consistent with the ionization cone, the *Chandra* images showed a flattened feature of hard continuum and Fe Kα circumnuclear emission with approximately 150 pc diameter, extending in the cross-cone direction. Although larger than what would be expected for the torus in this AGN, this feature could be part of a structure surrounding the torus. More recent deep observations of this structure (Marinucci et al. 2017) revealed clumpy emission in both the neutral and ionized Fe Kα lines.

Recent high resolution *Chandra* imaging of the Fe Kα emission of ESO 428-G014 (Fig. 7-20, left) provides another example of clumpy emission in the inner ~30 pc, which could be associated with the obscuring torus (Fabbiano et al. 2019). In the CT AGN NGC5643, a central flattened structure is visible with *Chandra* in the Fe Kα line (Fabbiano et al. 2018c), in the direction perpendicular to that of the ionization cone (Fig. 7-20, right). This structure is spatially coincident with the circumnuclear rotating molecular disk discovered with *ALMA* in this galaxy (Alonso-Herrero et al. 2018). The availability of both *ALMA* and *Chandra* observations gives us the first example in which both the obscuring structure ('torus') and the result of the nuclear X-rays interacting with this structure are directly imaged.

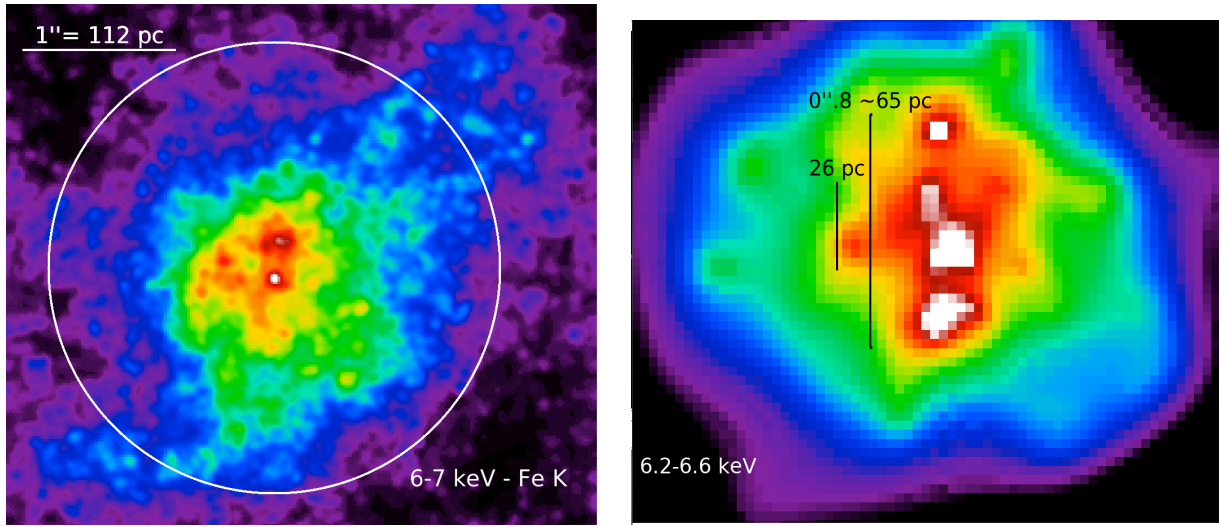

Fig. 7-20. Left, EMC2 reconstruction of the central region of ESO 428-G014 in the spectral band dominated by the Fe Kα emission observed with *Chandra,* showing both extended and clumpy emission (Fabbiano et al. 2019). Right, adaptive smoothing of the central north-south Fe Kα structure discovered with *Chandra* in NGC 5643 (Fabbiano et al. 2018c); this structure extends in the same direction as the CO (2-1) line emission and the central rotating CO (2-1) 26 pc disk discovered with *ALMA* (Alonso-Herrero et al. 2018).

**7.5.5 Is the Torus Porous?**

In the standard model, AGN photons escape along the axis of the torus and are collimated by the torus. However, at least in the case of NGC 4151 the presence of peculiar cloud velocities may suggest outflows in the cross-cone direction (Das et al. 2005). Moreover, the spectral variability of NGC 1365 (Risaliti et al. 2005) and NGC 4945 (Marinucci et al. 2012) shows that the obscuring structure at least in these two CT AGNs is composed of discrete clouds that at times move out of the AGN line of sight. However, the time variability implies cloud velocities consistent with the broad-line region, too fast for these clouds to be in torus. With *Chandra* imaging, X-ray emission has been detected in several AGNs in the direction perpendicular to the cone, where the torus should block their propagation. Examples are NGC 4151 (Wang et al. 2011c), Mkn 573 (Paggi et al. 2012), and ESO 428-G014 (Fabbiano et al. 2018a). These results suggest that the torus may be somewhat porous.

In ESO 428-G014 (Fabbiano et al. 2018a), the opening angle of the cone and the ratio of the X-ray emission detected in the cross-cone and cone regions imply that the cross-cone transmission of the obscuring torus is ~10% of that in the cone direction. In this CT AGN, the lack of energy dependence of the ratio of photons escaping from the cross-cone and cone regions would be in agreement with the partial obscuration picture, in which a fraction of photons escape in the cross-cone region with similar spectral distribution as those escaping in the cone region.

However, a different phenomenon may also contribute to the cross-cone X-ray emission. The interaction of radio jets with dense molecular disks may also produce cross-cone emission. Fabbiano et al. (2018b) note that in ESO 428-G014 the overall similarity of the morphology of the radio jet, the warm and the hot ISM are consistent with the predictions of recent 3D relativistic hydrodynamic simulations by Mukherjee et al. (2018) of the interaction of radio jets with a dense molecular disk. These simulations also predict a hot cocoon enveloping the interaction region, offering a possible alternative explanation to the presence of X-ray emission perpendicular from the bicone. All the CT AGNs where cross-cone extent has been reported so far host small radio jets.

Even while the interpretation of the *Chandra* results is still evolving, it is clear that these high angular resolution observations of AGNs are producing important new constraints on the AGN-galaxy interaction, and will also have repercussions for AGN modelling.

### 7.6 Looking Forward

The sub-arcsecond angular resolution and simultaneous spectral resolution available with *Chandra* ACIS have been proven a winning combination for the study of galaxies in X-rays. As discussed in this chapter, these unique *Chandra* capabilities have allowed the separation of point-like and extended emission in galaxies out to a few 10 Mpc away, resulting in the 'clean' observational characterization of these components, and resolving long-standing ambiguities from studies with inferior angular resolution. Comparison with similar angular resolution multi-wavelength data (*HST*, *Spitzer*, *JVLA* and *ALMA*) has further advanced our knowledge of the properties and evolution of galaxies and their active nuclei.

These near-universe observations have provided the baseline for studies of the evolution of galaxies and their X-ray emission components with redshift. Unique *Chandra* observational achievements include: (1) the study of populations of XRBs in a variety of stellar environments, and the realization that XRBs may be an important source of feedback in the early universe; (2) the physical and chemical characterization of the hot ISM, hot wind and halos, and their connection with star formation activity and galaxy mass; (3) the investigation of the full gamut of AGN activity, including the discovery of low-luminosity AGNs in dwarf galaxies both nearby and at higher redshift, which is consistent with the widespread presence of lower-mass nuclear black holes in these systems; and (4) the study of AGN –ISM interaction and feedback in both radio loud and radio faint AGNs.

While these results are impressive, looking forward it is also important to point out their limitations, resulting chiefly from the small collecting area of the *Chandra* telescope, which requires extremely long observations to achieve significant results. While a larger area X-ray telescope (*Athena*) is being developed in Europe, its angular resolution is several times worse than *Chandra*'s, making it unsuitable for future resolved studies of galaxies.

The *Lynx* mission instead, under study by NASA, is the natural telescope for progressing in this field. *Lynx*, with a collecting area similar to that projected for *Athena* (30-50 times larger than *Chandra*'s), seeks to maintain *Chandra*'s angular resolution, while providing substantially better spectral-imagers in both soft and harder energy ranges, a winning combination for the study of galaxies and their evolution.

AKNOWLEGEMENTS

At large part of this Chapter was written at the Aspen Center for Physics, and benefited from discussions with other visitors at the Center. The ACP is supported by National Science Foundation grant PHY-1607611. We thank Martin Elvis, Alessandro Paggi and Andreas Zezas for input and comments on the manuscript.